\documentclass[review]{elsarticle}

\usepackage{lineno,hyperref}
\usepackage{textcomp}
\usepackage{mathtools,amsmath, commath}
\usepackage{braket}
\usepackage{xcolor}
\modulolinenumbers[5]

\journal{Annals of Physics Special Issue: Philip W. Anderson}

\begin{document}

\begin{frontmatter}

\title{Spectral properties of three-dimensional Anderson model}




\author[ijsaddress]{J. \v Suntajs}
\author[fmfaddress]{T. Prosen}
\author[ijsaddress,fmfaddress]{L. Vidmar}


\address[ijsaddress]{Department of Theoretical Physics, J. Stefan Institute, SI-1000 Ljubljana, Slovenia}
\address[fmfaddress]{Department of Physics, Faculty of Mathematics and Physics, University of Ljubljana, SI-1000 Ljubljana, Slovenia}

\begin{abstract}
The three-dimensional Anderson model represents a paradigmatic model to understand the Anderson localization transition.
In this work we first review some key results obtained for this model in the past 50 years, and then study its properties from the perspective of modern numerical approaches.
Our main focus is on the quantitative comparison between the level sensitivity statistics and the level statistics.
While the former studies the sensitivity of Hamiltonian eigenlevels upon inserting a magnetic flux, the latter studies the properties of unperturbed eigenlevels.
We define two versions of dimensionless conductance, the first corresponding to the width of the level curvature distribution relative to the mean level spacing, and the second corresponding to the ratio of the Heisenberg time and the Thouless time obtained from the spectral form factor.
We show that both conductances look remarkably similar around the localization transition, in particular, they predict a nearly identical critical point consistent with other well-established measures of the transition.
We then study some further properties of those quantities:
for level curvatures, we discuss particular similarities and differences between the width of the level curvature distribution and the characteristic energy studied by Edwards and Thouless in their pioneering work [\href{https://doi.org/10.1088/0022-3719/5/8/007}{J.~Phys.~C.~{\bf 5},~807~(1972)}], in which the hopping at one lattice edge is changed from periodic to antiperiodic boundary conditions.
In the context of the spectral form factor, we show that at the critical point it enters a broad time-independent regime, in which its value is consistent with the level compressibility obtained from the level variance.
Finally, we test the scaling solution of the average level spacing ratio in the crossover regime using the cost function minimization approach introduced recently in
[\href{https://doi.org/10.1103/PhysRevB.102.064207}{Phys.~Rev.~B.~{\bf 102},~064207~(2020)}].
The latter approach seeks for the optimal scaling solution in the vicinity of the crossing point, while at the same time it allows for the crossing point to drift due to finite-size corrections.
We find that the extracted transition point and the scaling coefficient agree with those from the literature to high numerical accuracy.
\end{abstract}


\end{frontmatter}

\newpage
\section{Introduction}

The three-dimensional (3D) Anderson model is a paradigmatic model for the investigation of the Anderson localization transition occurring in disordered systems of non-interacting quantum particles. In an ideally ordered crystalline solid, electronic wave functions can be described as spatially extended Bloch waves due to translational symmetry. In most actual systems, however, such ideal crystallinity is only a crude approximation due to the almost inevitable presence of impurities, vacancies and dislocations, all of which constitute what will be termed as \emph{disorder} throughout this paper. As shown in 1958 in Philip W. Anderson's seminal paper, \emph{Absence of diffusion in certain random lattices}~\cite{Anderson_1958}, introducing random potentials to an otherwise ideal system can cause exponential localization of some of the system's electronic wave functions, provided that the randomness is sufficiently strong. Since electrons occupying localized states are restricted to finite spatial regions, they cannot contribute towards transport in the zero-temperature limit, when the effects of other degrees of freedom, such as phonons, are vanishingly small. Electrons in the extended states, on the other hand, usually contribute to a non-vanishing transport. Consequently, a disordered system is an insulator if only localized states exist below its Fermi energy and a metal if the Fermi level is in the region of the extended states~\cite{kramer_mackinnon_93}. Understanding the effects of disorder on localization of quantum mechanical wave functions is therefore of paramount importance in understanding the existence of metals and insulators and, in particular, the transitions between them. In 1977, less than 20 years after his ground-breaking paper, Philip W. Anderson was awarded with the Nobel Prize in Physics for his work on disordered systems~\cite{Lagendijk2009c}.

Soon after Anderson's original formulation of the problem in three dimensions, rigorous proofs~\cite{Mott1961, R.E.Borland1963} have been provided for the one-dimensional case, stating that all states become trivially localized for any finite disorder strength in one dimension. In three dimensions, as pointed out by Mott~\cite{Mott1968} in 1968, a critical value $W_\mathrm{c}$ exists for the disorder parameter, above which all of the system's eigenstates become exponentially localized. For sub-critical disorder parameter values, both localized and extended states are present in the system, separated by some critical energy $E_\mathrm{c},$ which has since been termed as the \emph{mobility edge}. Its existence is now recognized as a hallmark of a metal-insulator transition in a three-dimensional system. In the somewhat peculiar two-dimensional case, a rigorous proof establishing the existence of the localization transition is still lacking. It now appears that localization of all the states takes place in the thermodynamic limit for any finite disorder, much as in one dimension. Finite-size systems, on the other hand, might display metallic properties, thus rendering the numerical analysis of the problem particularly demanding. For a long while, much of our understanding of the localization phenomena in two dimensions stemmed from the findings of the \emph{scaling theory of localization}~\cite{Abrahams1979} which was put forth in 1979 by Abrahams, Anderson, Licciardello and Ramakrishnan. The scaling theory represented an important breakthrough as it provided an elementary description of the dimensionality's role in the onset of localization phenomena and it gave correct predictions about the transitions in one, two and three dimensions.
Within its framework, Anderson localization is described in the language of continuous quantum phase transitions and the scaling behaviour of the dimensionless conductance $g$ (to be defined in Sec.~\ref{sec:transition}) with the system size is considered. The choice of $g$ as the relevant variable for the scaling theory was particularly motivated by an earlier series of papers by Thouless and coauthors~\cite{edwards_thouless_72,Thouless1974a} from the mid-seventies. In those, they formulated a scaling description of the problem by establishing a relation between the sensitivity of the energy spectra to slight changes of the boundary conditions~\cite{edwards_thouless_72} while also backing-up their theoretical predictions with numerical results. The latter point is of particular importance for the present paper as well and we return to it in more detail in Sec.~\ref{sec:thouless}, where we follow their original calculations while making use of the modern hardware to study much larger system sizes than the ones available to the authors of the original study.

In the present paper, we focus on two cornerstones in studies of the Anderson localization transition.
The first corresponds to properties of the level sensitivity statistics, i.e., the response of Hamiltonian eigenlevels upon adding a magnetic flux.
One of the central objects in this context is the level curvature, which is related to the quantities studied by Thouless and coworkers in the seventies, and it also influenced the construction of the scaling theory of localization introduced above.
The second corresponds to the level statistics of unperturbed levels and is closely related to the findings of the \emph{random matrix theory} (RMT).
The relevance of the latter for the studies of the metal-insulator transitions were recognized already in the late eighties~\cite{altshuler_shklovskii_86, altshuler_zharekeshev_88}, when the changes in the statistical properties of the energy spectra of the Anderson Hamiltonian were studied across the transition. In accordance with the RMT, it was observed that the spectra exhibit repulsion between the nearest energy levels in the metallic regime while this feature is absent in the spectra of insulating systems.
The first accurate determinations of the transition point came in the early nineties~\cite{shklovskii_shapiro_93, hofstetter_schreiber_93}, followed by subsequent developments that allowed for the extraction of other scaling parameters~\cite{zharekeshev_kramer_95, zharekeshev_kramer_97}.

The paper is organised as follows: in Sec.~\ref{sec:transition} we introduce the 3D Anderson model and discuss different measures used in the investigation of the localization transition. We then proceed onto discussing our main results. In Sec.~\ref{sec:thouless}, we investigate two different forms of the dimensionless conductance $g$.
The first is related to the ratio of the width of the level curvature and the mean level spacing.
The second is related to the ratio of the Heisenberg and the Thouless time, which is extracted from the spectral form factor~\cite{mehta_91, suntajs_bonca_20a}.
The latter has been recently shown as a reliable probe for extracting the critical point in the 3D Anderson model~\cite{sierant_delande_20}.
We complement those studies by studying a related definition of conductance that resembles the implementation in the original paper by Edwards and Thouless~\cite{edwards_thouless_72}.
In Sec.~\ref{sec:scaling}, we use the cost function minimization approach, recently introduced in Ref.~\cite{suntajs_bonca_20b}, to obtain the scaling solutions of the mean ratio of the nearest level spacings $r$~\cite{Oganesyan2007a} across the transition. We verify that the obtained scaling parameters accurately agree with the ones extracted using different methods which are now widely accepted in the literature~\cite{slevin_ohtsuki_99, slevin_ohtsuki_14, slevin_ohtsuki_18}. 

\newpage
\section{Anderson localization transition: an overview} \label{sec:transition}

In this section, we provide a brief and incomplete overview of the studies of the 3D Anderson localization transition which have been conducted so far. At the same time, we prepare the groundwork for discussing our numerical results in later sections. We begin by defining the Anderson model Hamiltonian, then proceed onto discussing some commonly used methods for the detection of the transition and the current best estimates for the critical parameters of the transition. Particular focus is devoted to the studies of the spectral statistics, as those are closely related to the main results of our numerical work.

\subsection{The Anderson model}\label{sec:anderson_model}
The 3D Anderson model describes the motion of non-interacting spinless fermions on a cubic lattice in the presence of a disordered potential. Its Hamiltonian is defined as follows:
\begin{equation} \label{def_H}
\hat H = -t\sum_{\langle i,j\rangle} \left(\hat c_i^\dagger \hat c_j + {\rm h.c.}\right) + \frac{W}{2}\sum_{i=1}^V \epsilon_i \hat c_i^\dagger \hat c_i \,.
\end{equation}
Here, $\hat c_i^\dagger, \hat c_i$ are the fermionic creation and annihilation operators at the site $i$, respectively. The first term describes the nearest-neighbour hopping with $t$ being the corresponding kinetic energy scale, which we set equal to unity in all of our calculations. Randomness is introduced to the model through random on-site potentials $\varepsilon_i$ in the second term, where the values of $\varepsilon_i$ are i.i.d.~random variables uniformly distributed in the interval $[-1,1]$. The degree of disorder in the system is controlled by the disorder strength parameter $W.$ In terms of the basic symmetries, the Anderson Hamiltonian given by Eq.~\eqref{def_H} is time-reversal invariant and particle-hole symmetric.

Upon increasing the value of $W,$ the localization transition takes place at the critical value $W_c\approx16.5$~\cite{shklovskii_shapiro_93, slevin_ohtsuki_99}, separating the metallic and insulating phases. In the former, the electronic wave functions near the center of the band are extended over the whole volume of the lattice, while in the latter all wave functions are exponentially localized around some randomly distributed lattice sites. 
In what follows, we briefly review the studies of the transport properties and energy level statistics across the transition.
For a more comprehensive overview of the matter, we suggest, e.g., reviews in Refs.~\cite{lee_ramakrishnan_85, kramer_mackinnon_93, evers_mirlin_08, markos_06} for further reading.

\subsection{Measures of the transition} \label{sec:measures}
Since the original formulation of the problem, different approaches towards the investigation of the localization transition in the 3D Anderson model have been introduced.
An influential work was carried out by Thouless and coworkers in the seventies~\cite{edwards_thouless_72,Thouless1974a}, which established the connection of transport properties with the sensitivity of Hamiltonian eigenlevels with respect to a magnetic flux.
This subject is going to be elaborated on in more details in Secs.~\ref{sec:curvature} and~\ref{sec:edwards_thouless}.
Those works also inspired a theoretical framework of the transition put forth in 1979 with the scaling theory of localization~\cite{Abrahams1979}. In the latter, the dimensionless conductance $g$ of a $d$-dimensional hypercube of volume $L^d$ is considered as the scaling variable for the theory. For weak disorder in the metallic regime, the following relation between $g$ and conductivity $\sigma$ holds:
\begin{equation}\label{scaling_metallic}
    g(L)\frac{e^2}{\hbar} = \sigma L^{d-2}\,,
\end{equation}
where we set $e^2/\hbar \equiv 1$ further on.
For strong disorder, on the other hand, the system's wave functions are exponentially localized with the localization length $\xi$ and hence the conductance of a finite system is also exponentially vanishing as 
\begin{equation}\label{scaling_insulator}
    g\sim \exp(-L/\xi)\,.
\end{equation}
Introducing the logarithmic derivative of $g$ as
\begin{equation}\label{scaling_beta}
    \beta = \frac{\mathrm{d}\ln g}{\mathrm{d}\ln L}\,,
\end{equation}
we immediately see that $\beta=d-2$ in the metallic region, and hence $\beta>0$ for $d=3$. For strong disorder, on the other hand, one always has $\beta<0,$ indicating insulating behavior. The transition between a metal and an insulator occurs at the critical point $\beta(g_\mathrm{c})=0$ which can only happen in three (or more) dimensions where $\beta$ can assume both positive and negative values. In lower dimensional systems, the conductance always decreases with system size and so the critical point is never reached.

First numerical verification of the scaling theory came in the early eighties with the work of Kramer and MacKinnon~\cite{mackinnon_kramer_81, MacKinnon1983a}, and Pichard and Sarma~\cite{pichard_sarma_81a, pichard_sarma_81b}.
One of the most powerful methods to study the transition is the \emph{transfer matrix method} (TMM)~\cite{kramer_mackinnon_93}, which studies the scaling of the localization length $\xi$ (related to the smallest Lyapunov exponent) near the transition.
Using the TMM, the authors found the scaling behaviour predicted by the scaling theory and were able to accurately determine the critical scaling parameters~\cite{Kramer1990, mackinnon_94, Schreiber1996, slevin_ohtsuki_99, slevin_ohtsuki_18}.
Among those, we note the critical exponent $\nu$ describing the divergence of the localization length $\xi$ upon approaching the critical point:
\begin{equation}\label{eq:nu}
\xi = \left|W-W_\mathrm{c}\right|^{-\nu} \,.
\end{equation}
To date, the most accurate determination of $\nu$ has been performed using the massively parallel implementation of the TMM~\cite{slevin_ohtsuki_18}.
Currently accepted values for the two parameters are~\cite{slevin_ohtsuki_18}: 
\begin{equation} \label{def_Wc}
W_{\rm c} = 16.54, \hspace{5mm} \nu = 1.57 \,.
\end{equation}
In this paper, we use the above values of the critical parameters as the reference values.
The error bars of those critical parameters are expected to be small.
In particular, $W_{\rm c}$ from Eq.~(\ref{def_Wc}) is believed to be accurate within one percent, while $\nu$ is accurate within a few percents~\cite{slevin_ohtsuki_18}.

In the early 1990s, the connection between the Anderson localization transition and the findings of the RMT was explored.
Consequently, new approaches towards the investigation of the transition were introduced~\cite{shklovskii_shapiro_93, hofstetter_schreiber_93, zharekeshev_kramer_95, Varga1995a, Schreiber1996, zharekeshev_kramer_97, Zharekeshev1998}, which were based on the investigation of the Hamiltonian's spectral fluctuations rather than its wave functions. These new methods allowed for a surprisingly accurate determination of the transition point and its corresponding critical exponents~\cite{shklovskii_shapiro_93}. In accordance with the predictions of the RMT, various spectral observables display markedly different behavior on both sides of the localization transition. Within the framework of RMT, the onset of Anderson localization can be described as the transition between the Gaussian orthogonal ensemble (GOE) and the Poisson ensemble (PE) of random matrices. In the early works, the spectral observable of interest was typically the distribution of spacings between the nearest energy levels. In the GOE regime, as it turns out, it can be very well approximated by the Wigner surmise derived for two dimensional random matrices~\cite{mehta_91}
\begin{equation}\label{eq:wigner_surmise}
    P_\mathrm{GOE}(s) \cong \left(\frac{\pi}{2}\right)s\exp\left(-\frac{\pi}{4}s^2\right)\,,
\end{equation}
with $s$ being the nearest-level spacing. The distribution shows a linear level repulsion for small values of $s$ with a vanishing probability of degenerate energy levels. In the localized phase, where the spectral statistics are characterized by the PE, one instead has
\begin{equation}
    P_\mathrm{PE}(s) = \exp(-s)
\end{equation}
and thus no level repulsion is observed. In discrete spectra of finite-sized systems, one has
\begin{equation}
s_\alpha = E_{\alpha+1} - E_\alpha \geq 0\,,
\end{equation}
where $\{ E_\alpha \}$ denotes the ordered set of energy levels of the system. Due to a more convenient numerical implementation, an analogous quantity is used in more recent investigations, namely the ratio of the adjacent level spacings~\cite{Oganesyan2007a}:
\begin{equation}\label{eq:r_ratio}
    r_\alpha = \frac{\min\{s_\alpha, s_{\alpha+1}\}}{\max\{s_\alpha, s_{\alpha+1}\}}\,,\;\;\;\;\;
    r = \langle \, \langle r_\alpha\rangle_\alpha \rangle_W \,,
\end{equation}
where $\langle ... \rangle_\alpha$ denotes the average over $N_\alpha$ eigenstates in the center of the spectrum and $\langle ... \rangle_W$ denotes the average over $N_{\rm sample}$ different disorder configurations at a fixed $W$.
\begin{figure}[!t]
\includegraphics[width=1.0\columnwidth]{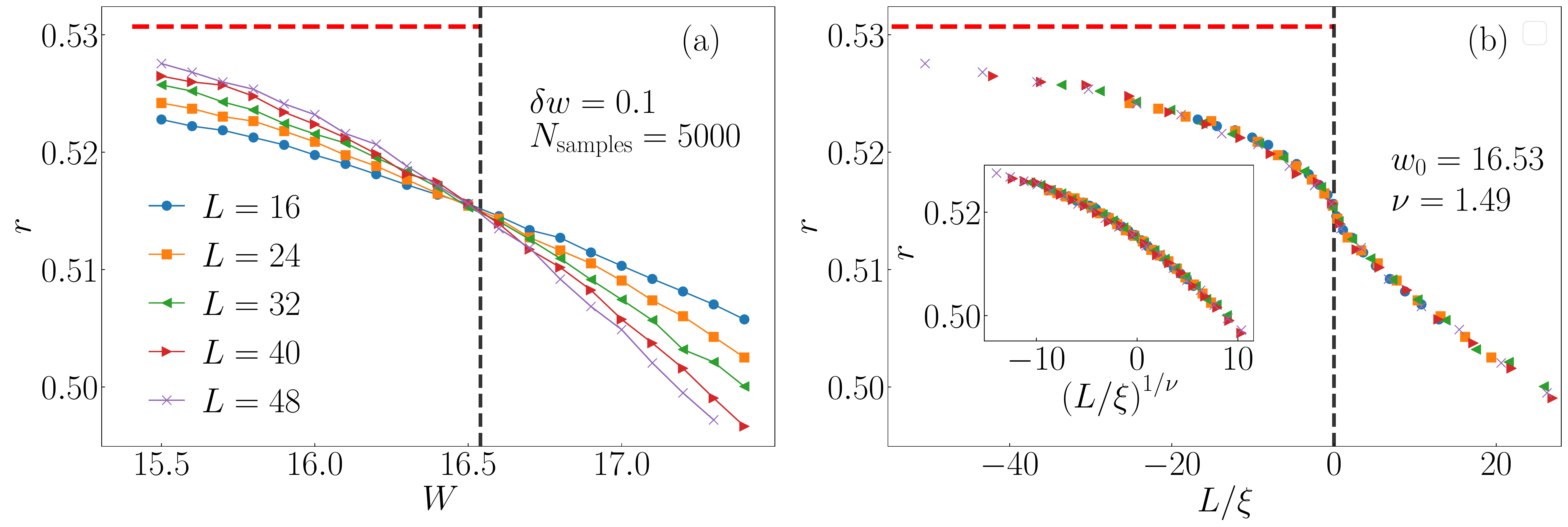}
\vspace{-0.1cm}
\caption{
Numerical scaling solution for the average level spacing ratio $r$ from Eq.~(\ref{eq:r_ratio}).
The averages are performed using $N_\alpha=500$ eigenstates and $N_\mathrm{samples}=5000$ different disorder realizations.
Horizontal dashed lines denote $r_{\rm GOE}$ and the vertical line in (a) denotes the critical disorder $W_{\rm c} = 16.54$.
(a) Dependence of $r$ on the disorder strength $W$ for different system sizes $L$.
We sampled the data points from the interval $\left[15.5, 17.4\right],$ using the step size of $\delta W=0.1$ between consecutive data points.
(b) The optimal scaling collapse obtained using the cost function minimization approach (see Sec.~\ref{sec:floating}) shows results as a function of $L/\xi$, where $\xi = {\rm sign}[W-w_0] \times |W-w_0|^{-\nu}$, yielding a perfect collapse.
We get the optimal parameters $w_0=16.53$ and $\nu=1.49$.
}
\label{fig_r_zoom}
\end{figure}
The distribution of $r_\alpha$ again displays universal behaviour in accordance with its underlying level statistics~\cite{lydzba_rigol_21, Atas2013}. In the extended and localized regimes, the average $r$ is expected to converge towards the GOE and PE results, respectively~\cite{Oganesyan2007a, Atas2013}. Tracing the change of the mean ratio $r$ from $r_\mathrm{GOE}\approx0.5307$ to $r_\mathrm{PE}\approx 0.3863$ allows one to distinguish the two regimes and perform a scaling analysis to obtain the critical parameter and scaling exponents. The quantity has only recently been employed in the studies of the 3D Anderson model~\cite{Tarquini2017b} and is also the subject of the present paper.
As an illustration, Figure~\ref{fig_r_zoom} shows the results of the scaling analysis performed using the \emph{cost function minimization} approach~\cite{suntajs_bonca_20b}. Using this method, 
we obtain the critical disorder value $w_0=16.53$ and an estimate for the critical exponent $\nu=1.49$, which are in good agreement with the values reported elsewhere in the literature and given by Eq.~\eqref{def_Wc}. We discuss the method and the more technical details of our calculations in Sec.~\ref{sec:scaling}.

Going beyond the spectral correlations between consecutive levels, we briefly discuss the \emph{spectral form factor} (SFF)~\cite{berry_85, mehta_91}, to which we return in significantly more detail in Sec.~\ref{sec:sff}. Much as the spectral statistics described above, the SFF also displays universal behaviour in the GOE and PE cases, however, its calculation also offers an insight into various time scales governing the dynamic behaviour of the studied systems. For instance, it allows for the determination of the Thouless time $t_\mathrm{Th}$ that denotes the time scale after which the quantum dynamics become universal and governed by the RMT.
Introducing the Heisenberg time $t_\mathrm{H}$ as the system's longest physically relevant time scale, inversely proportional to the mean level spacing $\Delta,$
one can express the dimensionless logarithmic conductance $g_{\rm SFF}$ as the logarithm of the ratio~\cite{suntajs_bonca_20a}
\begin{equation} \label{def_g_sff_sec2}
    g_{\rm SFF} = \log\frac{t_\mathrm{H}}{t_\mathrm{Th}} \,.
\end{equation}
For simplicity, we refer to the logarithmic conductance as conductance throughout the work.
It has recently been shown that requiring $t_\mathrm{H}/t_\mathrm{Th}=\mathrm{const}$ upon increasing the system size can be used as a very efficient tool for detection of the transition point in 3D and 5D Anderson models~\cite{sierant_delande_20}. Furthermore, calculating the conductance $g_{\rm SFF}$ as defined above, one can directly compare the results of a level statistics based approach with the results from level sensitivity statistics, such as the level curvatures.
We study the latter in Sec.~\ref{sec:curvature} and~\ref{sec:edwards_thouless}.

\begin{figure}[!t]
\includegraphics[width=1.0\columnwidth]{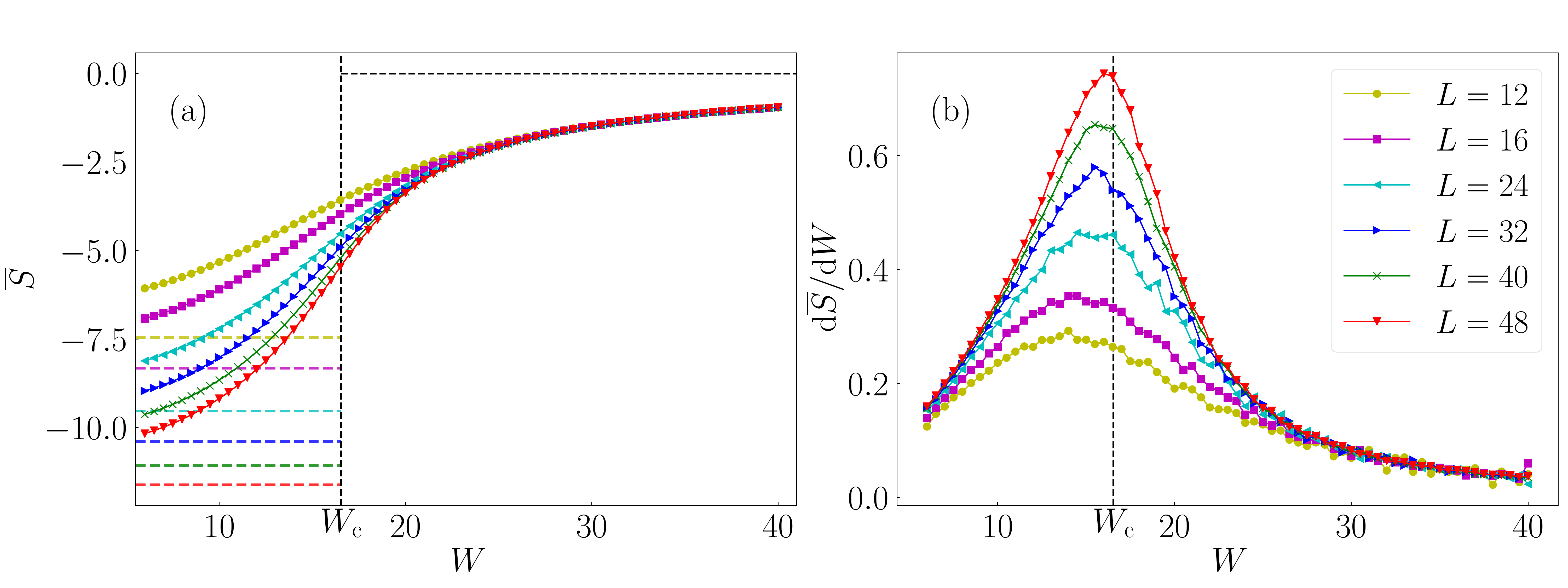}
\vspace{-0.1cm}
\caption{Analysis of the average participation entropy $\bar{S}$ from Eq.~(\ref{eq:participation_entropy}), averaged over $N_\alpha=500$ eigenstates in the center of the spectrum and $N_\mathrm{samples}=500$ different disorder realizations.
(a) $\bar{S}$ for different disorder strengths $W$ and system sizes $L$. Horizontal colored dashed lines denote the theoretical predictions for the extended states in the absence of disorder, $S_\mathrm{ext}=-d\ln L$. The opposite limit $S_\mathrm{loc}=0$, corresponding to ideally localized states, is denoted by the black horizontal dashed line. 
(b) The derivative of $\overline{S}$ with respect to disorder strength $W$ reveals a peak in the vicinity of the predicted transition point for the 3D Anderson model at $W_\mathrm{c}=16.54$. 
}
\label{fig1}
\end{figure}

Finally, we note that the Anderson localization transition can also be studied through the structure of Hamiltonian eigenfunctions.
Following the early works by Dean and Bell~\cite{Bell1970a}, this direction of research was pursued by many other authors (see, e.g., Ref.~\cite{evers_mirlin_08} for a review).
Even though the eigenfunction properties are not the focus of this work, we briefly comment on the analysis of the eigenfunction fluctuations using the average \emph{inverse participation ratio} (IPR).
For an eigenstate $\alpha$, the IPR is defined as follows:
\begin{equation}\label{eq:ipr}
P_\alpha^{-1} = \sum_j \left|\psi_{j,\alpha}\right|^4\,,
\end{equation}
where $\psi$ is the normalized wave function written in the site-occupational basis and $j$ labels sites on the Hamiltonian's underlying lattice.
An analogous quantity is often investigated, namely the \emph{participation entropy} $S$:
\begin{equation}\label{eq:participation_entropy}
S_\alpha = \ln P_\alpha^{-1}\,,\;\;\; \bar S = \langle \,\langle S_\alpha \rangle_\alpha \rangle_W\,,
\end{equation}
where the averages are defined analogously to those in Eq.~(\ref{eq:r_ratio}).
Plots of $\bar S$ as a function of the disorder strength and system size are shown in Fig.~\ref{fig1}(a).

In an ideally localized case, in which the probability density function equals unity on a single site and is zero elsewhere, it is immediately evident that $P^{-1}_\mathrm{loc.}=1$ and hence $\bar S = 0$.
This is the result at large $W$ in Fig.~\ref{fig1}(a).
In the opposite limit $W \to 0$, the wave functions are uniformly spread over the lattice volume and hence the wave function coefficients are of the order $L^{-d/2}$. Consequently, in the thermodynamic limit the IPR vanishes
as
$$
P^{-1}_\mathrm{ext} = \frac{C}{L^d},
$$
with the disorder-dependent prefactor $C$. In the absence of disorder, the wave functions are plane waves and hence $C=1,$ while for weak disorder deep in the extended regime $C$ assumes its GOE value of 3. 
The corresponding results for $\bar S$ in the former case are shown as horizontal dashed lines on the bottom of Fig.~\ref{fig1}(a).

The numerical results presented in Fig.~\ref{fig1}(a) are in agreement with the limiting cases discussed above.
However, it is less clear how to extract a quantitative prediction for the localization transition from those results.
Remarkably, as shown in Fig.~\ref{fig1}(b), a very accurate estimate of the critical point can be obtained when studying the behavior of $d\bar S/dW$ as a function of $W$.
It exhibits a peak at $W\approx W_c$, where $W_c$ from Eq.~(\ref{def_Wc}) is shown by a vertical dashed line in Fig.~\ref{fig1}(b).
The results show that the peak becomes sharper with increasing the system size, moreover, its position gets closer to $W_c$.

\subsection{What is the main goal of this paper?}
\label{sec:goal}

Having discussed some general features of the Anderson model, we now turn our attention to studies of some specific properties.
The main goal of this paper is to contribute to a unified understanding of the localization transition in the 3D Anderson model using different spectrum based techniques.
In Sec.~\ref{sec:thouless} we contrast two widely used approaches, the first is based on the level sensitivity statistics and the second is based on the level statistics, to quantitatively pinpoint the transition point.
In Sec.~\ref{sec:scaling} we test the quality of the numerical scaling solution across the transition using a recently introduced method~\cite{suntajs_bonca_20b}.
From the perspective of numerical calculations, we only report original results.
That said, several quantities studied here were studied before by other authors using identical or slightly modified definitions.
When applicable, we explicitly refer to those authors when discussing the results.

To obtain the energy spectra on which the subsequent analysis is performed, we employ the methods of exact numerical diagonalization. In Sec.~\ref{sec:thouless}, we use the method of full numerical diagonalization, since the knowledge of full spectra is needed for the calculation of quantities in question. To that end, we use the function \texttt{eigvalsh} from \texttt{numpy.linalg} \texttt{Python} library. Increased performance is achieved by using the version of \texttt{numpy} based on the Intel\textsuperscript{\textregistered} Math Kernel Library. In Sec.~\ref{sec:scaling}, we turn to partial diagonalization approaches, since the quantities introduced only require a portion of the system's eigenlevels for a successful calculation. To that end, we use the `shift-invert' approach described in Ref.~\cite{pietracaprina2018shift}.

\section{Indicators of localization transition: Level sensitivity statistics versus level statistics}
\label{sec:thouless}

We now proceed to the main part of this paper: the quantitative analysis of the determination of the critical point using two different approaches, i.e., the level sensitivity statistics and the level statistics.
In the first approach reported in Secs.~\ref{sec:curvature} and~\ref{sec:edwards_thouless} we study the sensitivity of Hamiltonian eigenlevels to a change of boundary conditions.
This approach was pioneered by Kohn in studies of ground state properties~\cite{kohn_64} and later generalized by Edwards and Thouless~\cite{edwards_thouless_72} to studies of arbitrary levels.
In the second approach reported in Sec.~\ref{sec:sff} we use the spectral form factor to extract the characteristic time scales of the system.

\subsection{Sensitivity to boundary conditions and level curvatures}
\label{sec:curvature}

To test the sensitivity of Hamiltonian eigenlevels towards the change of boundary conditions, we pierce the system by a flux $\Phi = L \phi$ along one of the axis (say, the $x$-axis) with $L$ sites.
The introduction of the flux modifies the kinetic energy term $\hat T \to \hat T(\phi)$ in the Anderson Hamiltonian, which is the first term in Eq.~(\ref{def_H}).
In fact, the new operator $\hat T(\phi)$ only modifies the hopping along the $x$-axis while other hopping terms remain unchanged.
Its contribution along the $x$-axis reads
\begin{equation} \label{def_T_phi}
    \hat T_{xx}(\phi) = -t \sum_{x=1}^L \sum_{\rho} \left( e^{i \phi} \hat c_{x+1,\rho}^\dagger \hat c_{x,\rho} + {\rm h.c.} \right) \,,
\end{equation}
where $\sum_\rho$ represents a double sum over all coordinates in the $\{y,z\}$ plane
and $\hat c_{L+1,\rho} = \hat c_{1,\rho}$.
We refer to $\hat T_{xx}(\phi=0)$ as $\hat T_{xx}$.
The sensitivity to boundary conditions is probed by considering a constant $\Phi \leq \pi$.
This implies $\phi \propto 1/L \ll 1$, and hence one can expand the Peierls phase in Eq.~(\ref{def_T_phi}) up to the second order in $\phi$, which yields
\begin{equation}
    \hat T_{xx}(\phi) = \hat T_{xx} - \phi \hat J_{xx} - \phi^2 \frac{\hat T_{xx}}{2} + {\cal O}(\phi^3)\,,
\end{equation}
where we introduced the particle current operator
$\hat J_{xx} = t \sum_{x=1}^L \sum_\rho \left( i \hat c_{x+1,\rho}^\dagger \hat c_{x,\rho} + {\rm h.c.} \right)$.
Using perturbation theory, the shift of the eigenlevels $E_\alpha$, expressed in the eigenbasis $\{ |\alpha\rangle \}$ of the Anderson Hamiltonian in Eq.~(\ref{def_H}), is
\begin{equation} \label{def_E_phi}
E_\alpha(\phi) = E_\alpha - \phi \langle \alpha | \hat J_{xx} | \alpha \rangle
- \phi^2 \left( \sum_{\beta\neq \alpha} \frac{|\langle\beta|\hat J_{xx}|\alpha\rangle|^2}{E_\beta - E_\alpha} + \frac{\langle\alpha| \hat T_{xx} |\alpha\rangle}{2} \right) + {\cal O}(\phi^3) \,.
\end{equation}
For the real eigenstates considered here, $\langle \alpha | \hat J_{xx} | \alpha \rangle = 0$.
The central object we investigate is the level curvature $c_\alpha$, defined as
\begin{equation} \label{def_c_alpha}
    E_\alpha(\Phi) = E_\alpha + \frac{1}{2}\Phi^2 c_\alpha\,,\;\;\;\;\;
    c_\alpha = \left( \frac{\partial^2 E_\alpha}{\partial \Phi^2} \right)_{\Phi=0}\;.
\end{equation}
In particular, the quantity that we numerically calculate in this section to get $c_\alpha$ is $[E_\alpha(\Phi) - E_\alpha] (2/\Phi^2)$ at $\Phi = 10^{-3}$.

Properties of the level curvatures as defined in Eq.~(\ref{def_c_alpha}) were subject of intense research in the past.
Inspired by the work by Edwards and Thouless~\cite{edwards_thouless_72}, perhaps the most widely studied property of $c_\alpha$ is the width of its distribution (more precisely, the mean of its absolute value), which is related to the system's conductivity via the Kubo-Greenwood formula~\cite{mott_70}.
This is going to be the main focus of our analysis below.
We note that subsequent work also devoted considerable attention to other properties of $c_\alpha$.
Examples include, e.g., the form of the distribution of $c_\alpha$~\cite{zakrzewski_delande_93, vonoppen_94, fyodorov_sommers_95} and the relation of level curvatures to other probes of level sensitivity such as the slopes of levels~\cite{wilkinson_88, akkermans_montambaux_92, simons_altshuler_93, braun_montambaux_94, fyodorov_sommers_95}.
In a related development, the thermal average of $c_\alpha$ was used as an indicator to distinguish quantum chaos from integrability in many-body systems~\cite{castella_zotos_95},
and the distribution of $c_\alpha$ was studied in the random-field Heisenberg models~\cite{filippone_brouwer_16, maksymov_sierant_19}.

The main focus in the study of level curvatures by Edwards and Thouless was devoted to the structure of the sum in the parenthesis of Eq.~(\ref{def_E_phi}).
Under the assumption that its main contribution stems from a few nearest levels for which the current matrix elements are sufficiently uncorrelated with the corresponding level spacings, the contribution of the {\it typical} term can be estimated from the Kubo-Greenwood formula~\cite{mott_70}.
In other words, one can estimate the width of the level curvature distribution $\langle |c| \rangle$ by a quantity that is proportional to the conductivity $\sigma$.
In the diffusive regime, the dependence of the system size then only emerges through the term $\phi \propto 1/L^2$ in Eq.~(\ref{def_E_phi}), which results in $\langle |c| \rangle \propto \sigma/L^2$.
We note that using the Einstein relation one can also view the latter proportionality as $\langle |c| \rangle \propto D/L^2$, where $D$ is the diffusion constant.

\begin{figure}[!t]
\includegraphics[width=1.0\columnwidth]{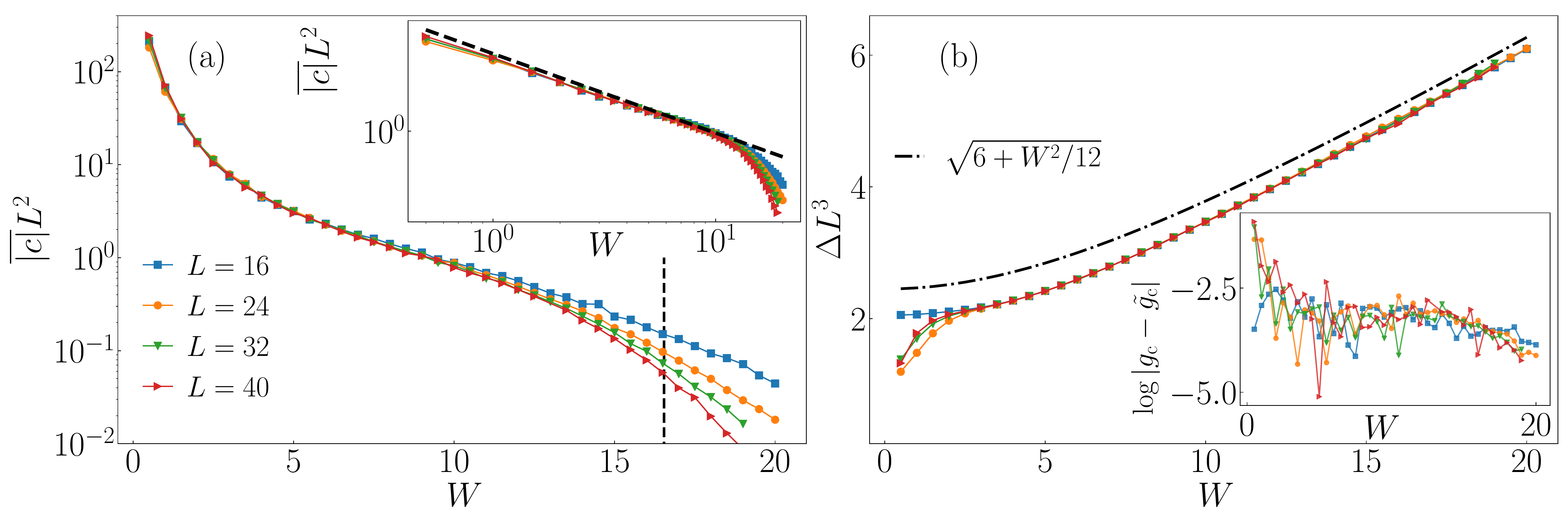}
\vspace{-0.1cm}
\caption{
(a)
Scaled curvature width $\overline{|c|}\, L^2$ as a function of disorder $W$.
The vertical dashed line denotes the critical disorder $W_c = 16.54$.
The inset shows the same result in a log-log scale.
The dashed line in the inset indicates the $\propto W^{-2}$ scaling in the small $W$ regime.
(b)
Scaled mean level spacing as a function of disorder $W$.
Symbols represent $\overline{\Delta}\, L^3$ and the dashed line is the scaled analytic result $\Delta\, L^3$ (see text for details).
The inset shows the difference $\log(|g_{\rm c} - \tilde g_{\rm c}|)$ of conductances $g_{\rm c}$ and $\tilde g_{\rm c}$ defined in Eqs.~(\ref{def_gc}) and~(\ref{def_gc_tilde}), respectively.
}
\label{fig_curvature}
\end{figure}

In this work we numerically study the width of the level curvature distribution in the diffusive regime as well as across the localization transition.
Specifically, we define the width of the curvature distribution as
\begin{equation} \label{def_c_over}
    \overline{|c|} = \langle \, \langle |c_\alpha| \rangle_\alpha \, \rangle_W \,,
\end{equation}
where $\langle ... \rangle_\alpha$ denotes the arithmetic mean over 500 eigenstates $\alpha$ around the center of the spectrum, and $\langle ... \rangle_W$ denotes the arithmetic mean over different disorder realizations for a given $W$.
As argued above, $\overline{|c|}$ is expected to scale as $\overline{|c|} \propto 1/L^2$ in the diffusive regime.
We verify this expectation in Fig.~\ref{fig_curvature}(a), where we plot the scaled curvature width $\overline{|c|}\, L^2$.
Indeed, the scaled results are $L$-independent in a wide regime of disorders that, in finite systems, terminates close to the critical point at $W_c$.
The inset of Fig.~\ref{fig_curvature}(a) shows the results in a log-log scale.
At small $W$, the results are in agreement with the scaling $\overline{|c|}\, L^2 \propto 1/W^2$, as predicted by perturbative arguments~\cite{montambaux_bouchiat_90}.

The results presented so far connect a particular property of level curvatures (specifically, the width of its distribution) to the conductivity $\sigma$.
However, as discussed in Sec.~\ref{sec:transition}, the central object in the scaling theory of localization is the conductance $g$.
These two quantities are in the diffusive regime related as $g = \sigma L^{d-2}$, where $d$ is the system dimension.
It is then suggested that one may define a conductance by multiplying the curvature width by a quantity that is proportional to $L^d$.
A natural candidate to achieve this is to divide it by the mean level spacing $\Delta$.
The conductance defined in this way is sometimes referred to as the Thouless conductance.

While the precise definition of the conductance is going to be introduced below, we first discuss the definition of the mean level spacing $\Delta$.
A convenient definition of $\Delta$ is through the ratio of the energy width $\Gamma_0$ and the Hilbert space dimension $V$, $\Delta = \Gamma_0/(\chi V)$, where the Hilbert space dimension is $V = L^3$ in the three dimensional system and $\chi=1/(2\sqrt{3})$ is a measure of the fraction of eigenstates (for a box distribution) in the interval between the mean energy $\bar E$ and $\bar E + \Gamma_0$.
In this way, $\Delta$ can be calculated analytically since the energy variance $\Gamma_0^2$ of the entire single-particle space is in the thermodynamic limit given by
\begin{equation} \label{def_variance}
\Gamma_0^2 = \frac{1}{V}{\rm Tr}\{\hat H^2\} - \left(\frac{1}{V}{\rm Tr}\{\hat H\}\right)^2
= \frac{W^2}{12} + 6 \,.
\end{equation}
Nevertheless, since the curvature width $\overline{|c|}$ is only averaged over the eigenstates in the center of the spectrum, we use an analogous definition to calculate the mean level spacing.
We hence calculate the latter numerically as the arithmetic average over those levels $\alpha$ that are also included in the calculation of $\langle|c_\alpha|\rangle_\alpha$ in Eq.~(\ref{def_c_over}).
We refer to it as $\langle\Delta_\alpha\rangle_\alpha$, and the average over disorder realizations,
\begin{equation}
    \overline{\Delta} = \langle\,\langle\Delta_\alpha\rangle_\alpha \,\rangle_W \,,
\end{equation}
is also defined analogously to Eq.~(\ref{def_c_over}).
Still, there is no substantial quantitative difference between $\Delta$ and $\overline{\Delta}$, as shown in the main panel of Fig.~\ref{fig_curvature}(b).

In the next step we define the conductance studied in this work.
As argued above, it corresponds to the ratio of the curvature width and the mean level spacing.
We define the conductance as a logarithm of this ratio to simplify its analysis for different regimes of disorder when it varies by several orders of magnitude.
In general, one can apply several different protocols of the disorder averaging.
Here we consider two such possibilities.
The first is the disorder average of the ratio
\begin{equation} \label{def_gc}
    g_{\rm c} = \log \left\langle \frac{\langle |c_\alpha|\rangle_\alpha}{\langle \Delta_\alpha \rangle_\alpha} \right\rangle_W \,,
\end{equation}
and the second is the ratio of the disorder averages,
\begin{equation} \label{def_gc_tilde}
    \tilde g_{\rm c} = \log  \frac{\langle \, \langle |c_\alpha|\rangle_\alpha \, \rangle_W}{\langle \, \langle \Delta_\alpha\rangle_\alpha \, \rangle_W} \,.
\end{equation}
The inset in Fig.~\ref{fig_curvature}(b) shows that the differences between $g_{\rm c}$ and $\tilde g_{\rm c}$ are actually very small for essentially all disorder values we are interested in.
We therefore proceed with studying $g_c$ from Eq.~(\ref{def_gc}) in all further calculations.
We attribute the origin of the numerical agreement between $g_{\rm c}$ and $\tilde g_{\rm c}$ to the fact that despite the average over eigenstates $\langle ... \rangle_\alpha$ includes a considerable amount of eigenstates (500), this number is still small when compared to the total number of eigenstates.
We note that 24 years ago, an analogous quantity was calculated by Braun, Hofstetter, MacKinnon and Montambaux for system volumes up to $V=12^3$ by averaging over roughly half of the spectrum~\cite{braun_hofstetter_97}.

\begin{figure}[!t]
\includegraphics[width=1.0\columnwidth]{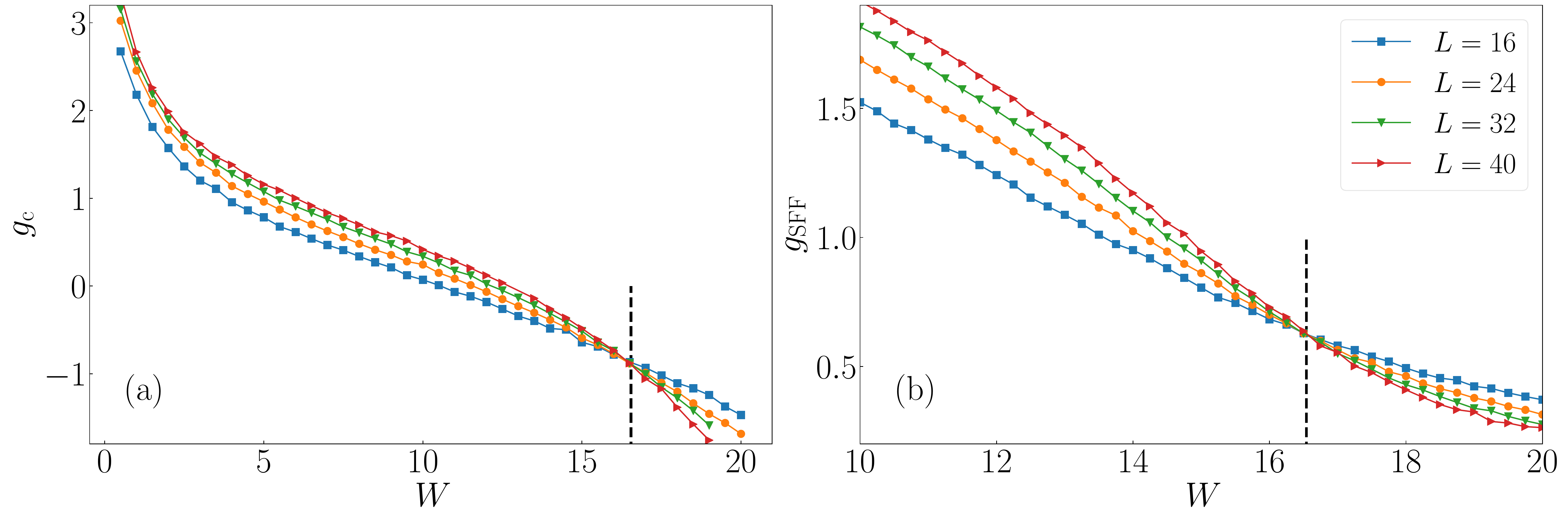}
\vspace{-0.1cm}
\caption{
(a) Conductance $g_{\rm c}$ as a function of disorder $W$.
We define $g_{\rm c}$ in Eq.~(\ref{def_gc}) through the width of the level curvature distributions.
(b) Conductance $g_{\rm SFF}$ as a function of disorder $W$.
We define $g_{\rm SFF}$ in Eq.~(\ref{def_g_sff}) through the Thouless time analysis from the spectral form factor.
Vertical dashed lines denote the critical disorder $W^* = 16.54$.
}
\label{fig_g}
\end{figure}

Figure~\ref{fig_g} represents one of the main results of this work.
In particular, Fig.~\ref{fig_g}(a) shows the conductance $g_c$ as defined in Eq.~(\ref{def_gc}).
In the diffusive regime at $W \ll W_c$, when the curvature width scales as $\langle|c_\alpha|\rangle_\alpha \propto 1/L^2$, the conductance increases as $g_{\rm c} \propto L$.
On the other hand, in the localized regime at $W \gg W_c$, the curvature width $\langle|c_\alpha|\rangle_\alpha$ decreases even faster with $L$ than the mean level spacing, and hence $g_{\rm c}$ decreases with increasing the system size.
Results in Fig.~\ref{fig_g}(a) suggest that there exists a disorder value $W$ for which $g_{\rm c}$ becomes independent of the system size.
In fact, this disorder value is indeed very close to the critical disorder $W_c$ obtained by using other measures of the transition discussed in Sec.~\ref{sec:transition}.

It is also remarkable that the results in Fig.~\ref{fig_g}(a) look very similar to the results in Fig.~\ref{fig_g}(b), in which we calculate another transition indicator, namely the conductance $g_{\rm SFF}$, which was introduced in Sec.~\ref{sec:measures}.
The latter is obtained from the level statistics analysis of the Hamiltonian without the flux.
Further details about $g_{\rm SFF}$ are discussed in Sec.~\ref{sec:sff}.

\begin{figure}[!t]
\includegraphics[width=1.0\columnwidth]{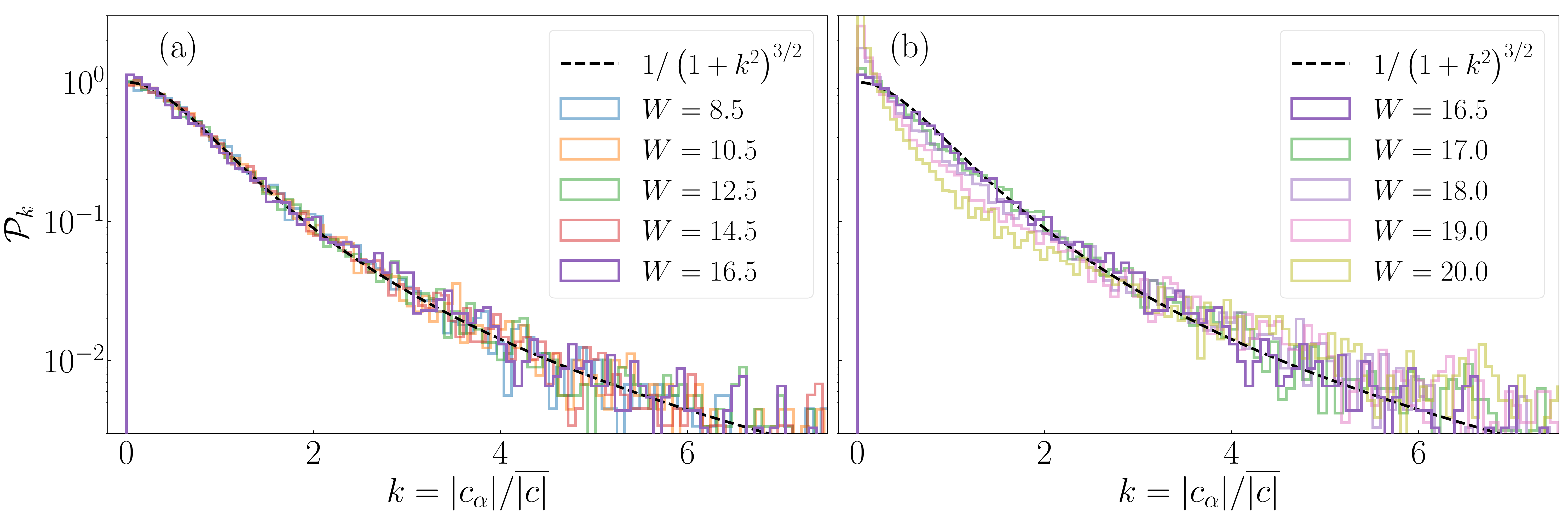}
\vspace{-0.1cm}
\caption{Distributions of level curvatures $\left|c_\alpha\right|$ for different disorder strengths $W$ and the largest available system size $L=40$. For each disorder realization, we chose $N_\alpha=500$ level curvatures corresponding to states near the center of the energy spectrum. Using $N_\mathrm{samples}=22$ different disorder realizations, 11000 curvature values thus appear in the histogram binning. To allow for comparison of distributions for different disorder parameter values, we study the distribution ${\cal P}_k$ of scaled curvatures $k = |c_\alpha|/\overline{|c|}$.
(a) Scaled curvature distributions for different disorder strengths in the delocalized regime for subcritical disorder strengths, $W\leq W_{\rm c} = 16.54$.
The numerical results match with the analytical prediction $P_1(k)$ given by Eq.~\eqref{eq:von_oppen_dist}.
(b) Numerical results for the scaled curvature distributions in the localized regime $W > W_{\rm c}$ indicate a departure from the universal behavior shown in panel (a).
}
\label{fig_kohn_distributions_semilog}
\end{figure}

In addition to the studies of the mean level curvatures, Fig.~\ref{fig_kohn_distributions_semilog} displays the properties of the level curvature distributions upon changing the disorder strength parameter. Once the level curvatures $|c_\alpha|$ are rescaled by the mean level curvature $\overline{|c_\alpha|}$, distributions in the delocalized regime exhibit universal properties~\cite{zakrzewski_delande_93, vonoppen_94, braun_hofstetter_97}. First empirically guessed by Zakrzewski and Delande~\cite{zakrzewski_delande_93} and then analytically confirmed by von Oppen~\cite{vonoppen_94}, the distribution of level curvatures is given by the expression
\begin{equation}\label{eq:von_oppen_dist}
    P_\beta(c)=\frac{C_\beta}{\left(1+c^2\right)^{(\beta+2)/2}},
\end{equation}
where $C_\beta$ is the normalizing constant with $\beta=1$ and $C_1=1$ in the GOE case with time-reversal symmetry (our interest here) and $\beta=2$ in the GUE case without time-reversal symmetry.
Our numerical results in Fig.~\ref{fig_kohn_distributions_semilog}(a) for the largest system size available, $L=40$, seem to confirm the validity of the universal prediction given by Eq.~\eqref{eq:von_oppen_dist} up to the critical disorder strength $W_{\rm c}$.
Deviations from the universal prediction become pronounced when the disorder is increased above $W_{\rm c}$, as shown in Fig.~\ref{fig_kohn_distributions_semilog}(b).
The distributions in the localized regime develop a sharper peak at smaller curvatures and broader tails, consistent with previous studies (see, e.g., Ref.~\cite{braun_hofstetter_97}).

\subsection{Relation to the numerical implementation by Edwards and Thouless}
\label{sec:edwards_thouless}
We extend the previous section by making an explicit connection to the numerical implementation of the curvature calculation by Edwards and Thouless in their pioneering work~\cite{edwards_thouless_72}.
They calculated the change of eigenenergies upon modifying the boundary conditions from periodic to antiperiodic at one lattice edge.
Within the formalism in Sec.~\ref{sec:curvature}, this corresponds to the introduction of the flux $\Phi = \pi$ along the $x$-axis.
We define the change of energy of an eigenlevel $\alpha$ as
\begin{equation} \label{eq:delta_e_thouless}
    \delta E^{(\pi)}_{\alpha} = E_\alpha(\Phi=\pi) - E_\alpha \,.
\end{equation}
Then, we define the Edwards-Thouless (ET) energy as
\begin{equation} \label{def_E_ET}
    E_{\rm ET} = \langle \, \langle |\delta E^{(\pi)}_\alpha |\rangle_\alpha \, \rangle_W \,,
\end{equation}
where $\langle ... \rangle_\alpha$ denotes the arithmetic mean over 500 eigenstates $\alpha$ around the center of the spectrum, and $\langle ... \rangle_W$ denotes the arithmetic mean over different disorder realizations for a given $W$.
The definition of $E_{\rm ET}$ in Eq.~(\ref{def_E_ET}) is, up to a normalization constant, analogous to the average over the absolute values of curvatures at $\Phi \ll 1$ defined in Eq.~(\ref{def_c_over}).

\begin{figure}[!t]
\includegraphics[width=1.0\columnwidth]{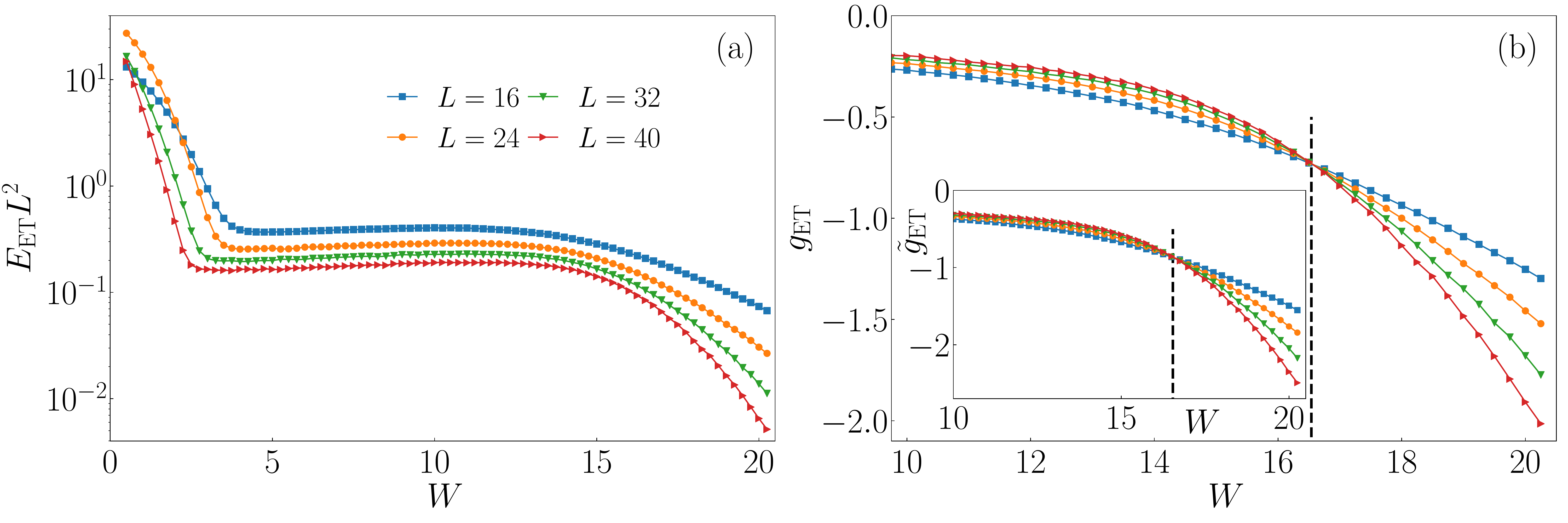}
\vspace{-0.1cm}
\caption{
(a)
The scaled Edwards-Thouless energy $E_{\rm ET} L^2$ as a function of disorder $W$ at different system sizes $L$.
(b)
Scaling of the conductance $g_{\rm ET}$, as defined in Eq.~(\ref{def_g_et}), across the localization transition.
The inset shows the conductance $\tilde g_{\rm ET}$, in which the only difference with respect to $g_{\rm ET}$ is that the arithmetic mean over eigenlevels is replaced by a geometric mean.
Vertical dashed lines denote the critical disorder $W_{\rm c} = 16.54$.
}
\label{fig_EdwardsThouless}
\end{figure}

While Edwards and Thouless studied the Anderson model on a square lattice and a diamond lattice~\cite{edwards_thouless_72}, we here implement an analogous numerical procedure for the cubic lattice.
The first question that we address is the system size dependence of the Edwards-Thouless energy $E_{\rm ET}$.
For the width of the curvature distribution,
Fig.~\ref{fig_curvature}(a) demonstrated that $\overline{|c_\alpha|} \propto L^{-2}$ in the diffusive regime.
In contrast,  Fig.~\ref{fig_EdwardsThouless}(a) shows that the scaled Edwards-Thouless energy $E_{\rm ET} L^2$ is, for the system sizes under investigation, not a system size independent quantity.
This may suggest that higher order terms in Eq.~(\ref{def_E_phi}) may contribute to an additional system size dependence that is absent in the case when $\Phi \ll 1$.
This also precludes one to make an explicit connection between the scaled energy $E_{\rm ET} L^2$ and the conductivity.

However, we show below that $E_{\rm ET}$ still represents a useful quantity to pinpoint the localization transition in finite systems.
To this end we define the logarithm of the ratio of the Edwards-Thouless energy and the mean level spacing,
\begin{equation} \label{def_g_et}
    g_{\rm ET} = \log \left\langle \frac{\langle |\delta E^{(\pi)}_\alpha |\rangle_\alpha}{\langle \Delta_\alpha \rangle_\alpha} \right\rangle_W\,.
\end{equation}
This quantity can be seen as the analog of the conductance $g_{\rm c}$ from Eq.~(\ref{def_gc}).
Results for $g_{\rm ET}$ versus $W$ are shown in the main panel of Fig.~\ref{fig_EdwardsThouless}(b).
Their main feature is that the crossing point of $g_{\rm ET}(W,L)$, i.e., the disorder value at which $g_{\rm ET}$ becomes system size independent, is indeed very close to the critical disorder $W_c$.
In this spirit, one can apply $g_{\rm ET}$ to pinpoint the localization transition with similar numerical accuracy as compared to the conductances $g_{\rm c}$ and $g_{\rm SFF}$ shown in Fig.~\ref{fig_g}.

We finally comment on the impact of averaging over the system's eigenstates.
In all the numerical calculations so far we performed the arithmetic mean over eigenstates.
We also tested the robustness of the results upon replacing the arithmetic mean with the geometric mean.
We show an explicit example for $g_{\rm ET}$ in Eq.~(\ref{def_g_et}), in which the arithmetic mean $\langle ... \rangle_W$ is replaced by the geometric mean, yielding $g_{\rm ET} \to \tilde g_{\rm ET}$.
Results for $\tilde g_{\rm ET}$ are shown in the inset of Fig.~\ref{fig_EdwardsThouless}(b).
They show that also in this case, the disorder value at which $\tilde g_{\rm ET}$ is independent of the system size is very close to the critical disorder obtained by other measures of the transition.
Hence, these results suggest that the type of averaging does not lead to any significant quantitative difference with respect to the identification of the critical point.
Similarly, we also observed (not shown) that the type of the disorder averaging does not have any significant impact on the scaling properties of $g_{\rm c}$ presented in Fig.~\ref{fig_curvature}(a).

\subsection{Spectral form factor} \label{sec:sff}

Results from the previous sections are now contrasted to the analysis based on spectral statistics of the Anderson Hamiltonian $\hat H$ without the presence of a flux.
The central quantity in the analysis is the spectral form factor (SFF), which is a Fourier transform of the spectral two-point correlations, defined as 
\begin{equation} \label{def_Kt}
K(\tau) = \frac{1}{Z} \left\langle \left|\sum_{\alpha = 1}^V \rho(\varepsilon_\alpha) e^{-i 2\pi\varepsilon_\alpha \tau}\right|^2 \right\rangle_W \, .
\end{equation}
In this expression, $\{ \varepsilon_\alpha \}$ denote the complete ordered set of single-particle eigenvalues $\{ E_\alpha \}$ of $\hat H$ after spectral unfolding.
The unfolding procedure is initiated by defining the cumulative spectral function ${\cal G}(E) = \sum_\alpha \Theta(E-E_\alpha)$, where $\Theta$ is the unit step function, followed by fitting a polynomial $\bar g_3(E)$ of degree 3 to ${\cal G}(E)$.
The unfolded eigenvalues are then defined as $\varepsilon_\alpha = \bar g_3(E_\alpha)$.
We thereby set the local mean level spacing to unity at all energy densities, and we refer to the corresponding time $\tau$ in these units as the scaled time.
In our analysis we consider properties of the Heisenberg time and the Thouless time.
The Heisenberg time is defined as the inverse mean level spacing (times $\hbar$, which is set to unity), and the Thouless time is defined below.
Since we are only interested in the ratio of those times, it is sufficient to consider them in their scaled units.
As a consequence of the unfolding procedure, the scaled Heisenberg time is $\tau_{\rm H} = 1$.

Note that in the definition of the SFF $K(\tau)$ in Eq.~(\ref{def_Kt}), we included the Gaussian filter $\rho(\varepsilon_\alpha)$, which eliminates contributions from the spectral edges.
The filtering function is defined as $\rho(\varepsilon_\alpha) = \exp\{-\frac{(\varepsilon_\alpha - \bar\varepsilon)^2}{2\eta\Gamma^2}\}$,
where $\bar\varepsilon$ and $\Gamma^2$ are the mean energy and the variance, respectively, for a given disorder realization, and $\eta=0.5$ controls the effective fraction of eigenstates included in $K(\tau)$.
The normalization $Z = \langle \sum_\alpha |\rho(\varepsilon_\alpha)|^2 \rangle_W$ then assures $K(\tau\gg 1) \simeq 1$.
Since the SFF is a not a self-averaging quantity~\cite{Cotler2017}, an additional averaging needs to be performed in order to obtain physically relevant results. As suggested in the literature~\cite{Cotler2017}, we use both ensemble and time averaging to reduce the fluctuations of the SFF. We perform the former by averaging our results over different disorder realizations at a given $W$, which we denote by $\langle ... \rangle_W$. Using a sufficiently narrow sliding window~\cite{Cotler2017}, we then perform the time averaging as well. 

\begin{figure}[!t]
\includegraphics[width=1.0\columnwidth]{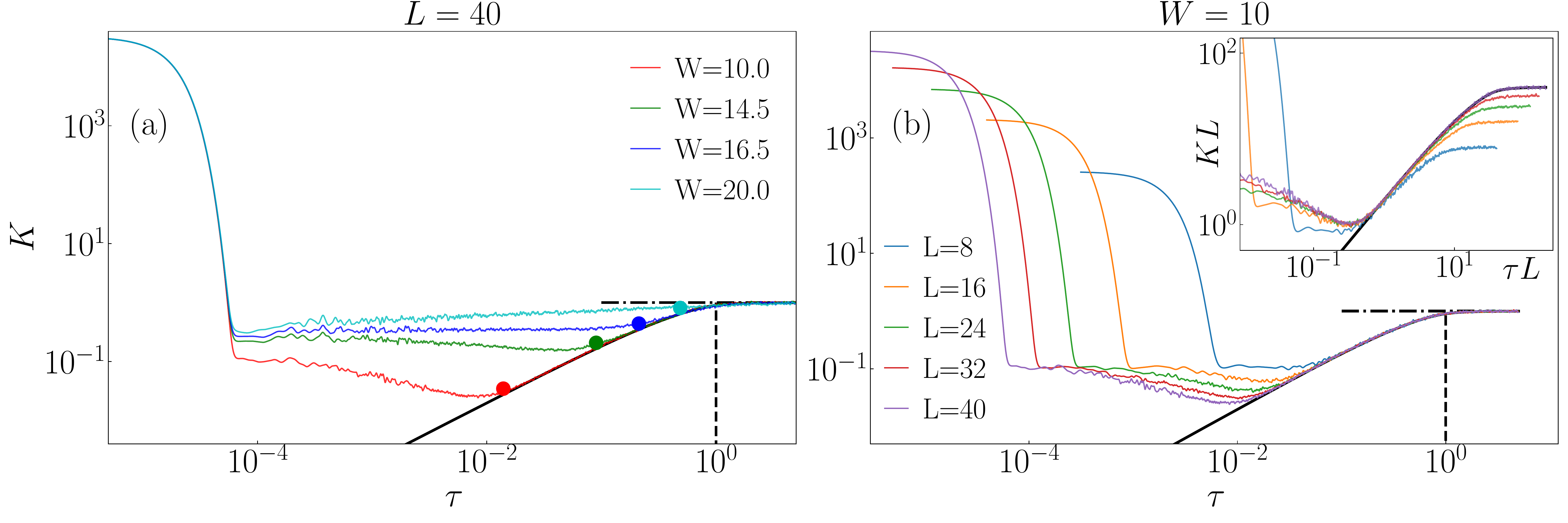}
\vspace{-0.1cm}
\caption{
(a)
The SFF $K(\tau)$ at a fixed system size with $L=40$ and different disorders $W$ that correspond to the values below, at, and above the localization transition.
Circles denote the values of the scaled Thouless time $\tau_{\rm Th}$, which are determined as described in the main text.
(b)
The SFF $K(\tau)$ at a fixed disorder $W=10$ and different system sizes $L$.
The inset shows the rescaled SFF $K(\tau) L$ versus $\tau L$, which demonstrates the $L^2$ dependence of the Thouless time in physical units (units of hopping).
In both panels, the thick black solid line is the GOE results $K_{\rm GOE}(\tau)$ from Eq.~(\ref{def_K_goe}) and the vertical dashed line denotes the scaled Heisenberg time $\tau_{\rm H} = 1$.
}
\label{fig_sff1}
\end{figure}

In physical systems that exhibit quantum chaos, the SFF $K(\tau)$ is usually nonuniversal at short times, apart from the ultrashort time $\tau \ll 1$ when $K(\tau)\gg 1$ (it is dominated by the density of states or the filter function $\rho$), which is not the focus of this work.
Our main interest is the long time regime, in which $K(\tau)$ becomes universal in the sense that it matches the prediction from the Gaussian orthogonal ensemble (GOE) of the RMT~\cite{mehta_91},
\begin{equation} \label{def_K_goe}
    K_{\rm GOE}(\tau) = 2\tau - \tau \ln(1+2\tau)\,.
\end{equation}
As a consequence, there exists a characteristic time which marks the crossover from a nonuniversal short time dynamics to the universal long time dynamics.
This time is now denoted as the Thouless time in the literature, even though Thouless himself, as argued in Secs.~\ref{sec:curvature} and~\ref{sec:edwards_thouless}, was primarily interested in the level sensitivity statistics rather than level statistics.
Numerically, we determine the scaled Thouless time $\tau_{\rm Th}$ from the criterion $|\log[K(\tau_{\rm Th})/K_{\rm GOE}(\tau_{\rm Th})]| = \epsilon$, as described in~\cite{suntajs_bonca_20a}.
We set $\epsilon = 0.08$ to get the results for $\tau_{\rm Th}$, which are shown in Figs.~\ref{fig_g}(b) and~\ref{fig_sff1}(a).

The single-particle spectrum of the 3D Anderson model below the localization transition is expected to comply with the GOE predictions~\cite{altshuler_shklovskii_86, altshuler_zharekeshev_88}.
In the context of the SFF $K(\tau)$, this means that there exists a regime at $\tau_{\rm Th} < \tau<1$ in which $K(\tau) \approx K_{\rm GOE}(\tau)$, as discussed above.
This is illustrated in Fig.~\ref{fig_sff1}(a) for a lattice with $V=40^3 = 64000$ sites.
Below the localization transition, i.e., at $W=10$ and 14.5 in Fig.~\ref{fig_sff1}(a), the results show that $\tau_{\rm Th}$, denoted by circles, is much smaller than 1.
In contrast, above the localization transition, i.e., at $W=20$ in Fig.~\ref{fig_sff1}(a), there is essentially no regime at $\tau < 1$ in which $K(\tau) \approx K_{\rm GOE}(\tau)$, and hence our numerical analysis yields $\tau_{\rm Th} \approx 1$.

Some properties of the SFF in the 3D Anderson model have been recently studied by Sierant, Delande and Zakrzewski~\cite{sierant_delande_20}.
They showed that in the diffusive regime well below the critical point, the Thouless time in physical units scales with the linear system size as $\propto L^2$.
We here corroborate this result with the analysis in Fig.~\ref{fig_sff1}(b) for the particular disorder value $W=10$.
The results in the main panel of Fig.~\ref{fig_sff1}(b) show a drift of the scaled Thouless time $\tau_{\rm Th} \to 0$ with increasing the system size, which suggests that the Thouless time in physical units scales slower than $\propto L^3$.
In the inset of Fig.~\ref{fig_sff1}(b), we plot the scaled SFF as a function of $\tau L$.
In this case the scaled Thouless time appears to be independent of the system size, which indicates that the Thouless time in physical units is proportional to $L^2$.
We also note that the authors of Ref.~\cite{sierant_delande_20} also studied the Thouless time in physical units divided by the system volume, which is related to the conductance $g_{\rm SFF}$ that we define below.

In Eq.~(\ref{def_gc}) of Sec.~\ref{sec:curvature} we introduced the conductance $g_{\rm c}$ as an indicator of the localization transition based on the level sensitivity statistics.
Here we introduce an analogous quantity (also termed conductance) based on the SFF analysis.
In particular, the indicator of the transition is defined as the logarithm of the ratio of the Heisenberg and the Thouless time~\cite{suntajs_bonca_20a},
\begin{equation} \label{def_g_sff}
    g_{\rm SFF} = \log \left( \frac{\tau_{\rm H}}{\tau_{\rm Th}} \right) = - \log \tau_{\rm Th} \,,
\end{equation}
where the ratio has been expressed in terms of scaled times, and hence the scaled Heisenberg time $\tau_{\rm H}$ can be set to unity.
The definition of $g_{\rm SFF}$ in Eq.~(\ref{def_g_sff}) using the ratio of scaled times is identical to the one in Eq.~(\ref{def_g_sff_sec2}) that uses the ratio of times in physical units.
Results for the conductance $g_{\rm SFF}$ as a function of $W$ are shown in Fig.~\ref{fig_g}(b).
They share striking similarities with the conductance $g_{\rm c}$ studied in Sec.~\ref{sec:curvature}.
In particular, they both provide an efficient measure to pinpoint the transition point since in both cases the curves for different system sizes cross at the disorder value that is extremely close to the well-established critical disorder value discussed in Sec.~\ref{sec:transition}.

To summarize our results for the scaling of conductances across the transition, they suggest that a very accurate estimate of the localization transition is obtained by requiring
\begin{equation} \label{def_criterion}
    \lim_{L\to\infty} g_\mu^* = {\rm const.}\,,
\end{equation}
where $\mu = $ c, ET, or SFF [cf.~Eqs.~(\ref{def_gc}), (\ref{def_g_et}) and~(\ref{def_g_sff}), respectively].
This quantitative agreement signals a connection between the system's sensitivity to boundary conditions and the validity of the underlying RMT description for unperturbed levels.

\begin{figure}[!t]
\includegraphics[width=1.0\columnwidth]{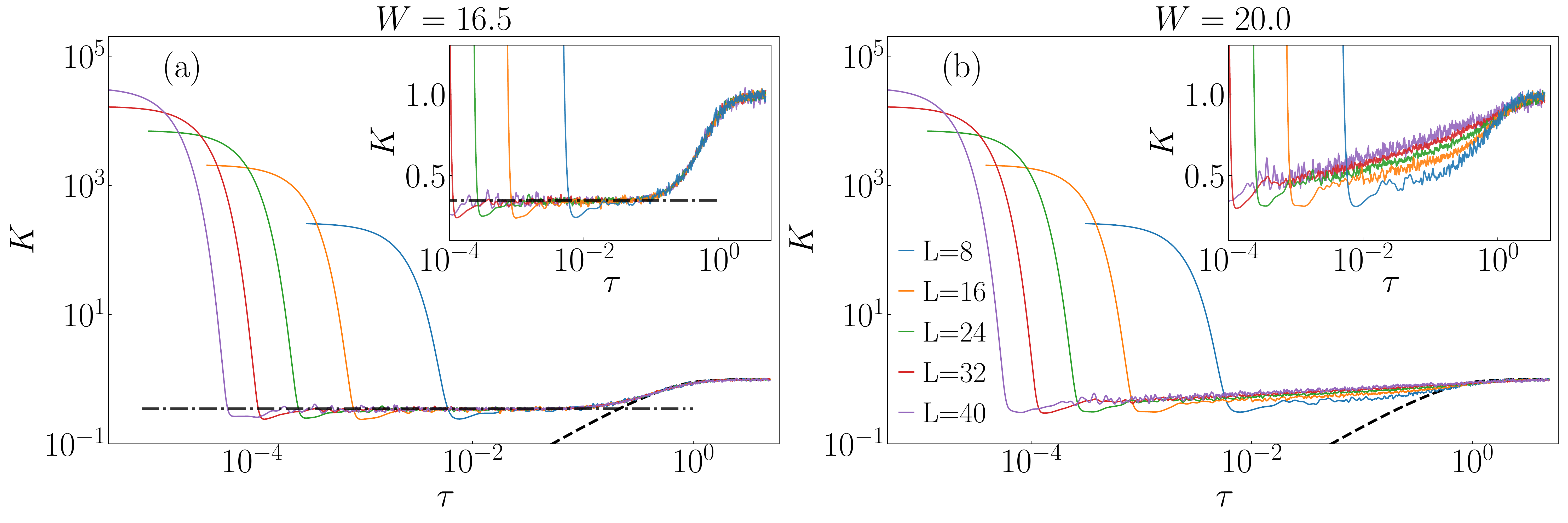}
\vspace{-0.1cm}
\caption{
Comparison of the SFF $K(\tau)$ (a) at the critical point $W=16.5 \approx W_{\rm c}$, and (b) in the localized regime at $W=20$.
Results in the insets show identical data as in the main panels, but on a log-linear scale.
Dashed lines are the GOE predictions $K_{\rm GOE}(\tau)$ from Eq.~(\ref{def_K_goe}).
Horizontal dashed-dotted line in (a) is a fit of the $L=40$ results in the interval $\tau \in [10^{-3},10^{-1}/2]$ to a constant, yielding $\kappa \approx 0.35.$}
\label{fig_sff2}
\end{figure}

We next consider some further details of the SFF $K(\tau)$ in the vicinity of the critical point.
Equation~(\ref{def_criterion}) suggests that $\tau_{\rm Th}^*$ does not scale with the system size at the critical point $W_c$.
This is indeed consistent with the results in Fig.~\ref{fig_sff2}(a) for $W=16.5 \approx W_c$, which shows $K(\tau)$ for different system sizes.
Moreover, the results in Fig.~\ref{fig_sff2}(a) also suggest two related observations, which we discuss below.

The first observation is that $\tau_{\rm Th}^* < 1$.
This suggests the emergence of a time interval $\tau \in [\tau_{\rm Th}^*, 1]$, in which the SFF $K(\tau)$ at the critical point matches the GOE prediction $K_{\rm GOE}(\tau)$.
Hence, within the framework of the SFF analysis introduced here, one may interpret the results as the Thouless time being smaller than the Heisenberg time at the localization transition.

The second observation is that the SFF $K(\tau)$ at $\tau \ll 1$ is, to a high numerical accuracy, independent of $\tau$ for several orders of magnitude.
The value of $K(\tau)$ in this regime is marked by a horizontal dashed-dotted line in Fig.~\ref{fig_sff2}(a).
We note that in the localized regime at $W > W_c$ we do not observe a similar structure of $K(\tau)$, at least for the system sizes under investigation. 
This is illustrated for $W=20$ in Fig.~\ref{fig_sff2}(b).
In particular, the inset of Fig.~\ref{fig_sff2}(b) shows a different finite-size trend of the results at $W=20$ when compared to those at the critical point.
Results for $K(\tau)$ at the critical point appear to be consistent with other measures of level statistics such as the level variance, which was shown to scale as $\Sigma^2(\delta E) = \kappa \, \delta E$ at the critical point, where $\kappa$ is the level compressibility $0 < \kappa < 1$~\cite{altshuler_zharekeshev_88, braun_montambaux_95, zharekeshev_kramer_95, chalker_kravtsov_96, braun_montambaux_98, ndawana_roemer_02}.
Indeed, if the latter expression holds true at all energy scales, it corresponds to a constant SFF $K(\tau) = \kappa$, and can be traced back to multifractality of eigenfunctions~\cite{chalker_kravtsov_96}.
The numerical value of the plateau height $\kappa = 0.35$ extracted from Fig.~\ref{fig_sff2}(a) is not far from the value $\kappa \approx 0.25$ obtained in the study of the level variance by Altshuler, Zharekeshev, Kotochigova and Shklovskii~\cite{altshuler_zharekeshev_88}.

\begin{figure}[!t]
\includegraphics[width=1.0\columnwidth]{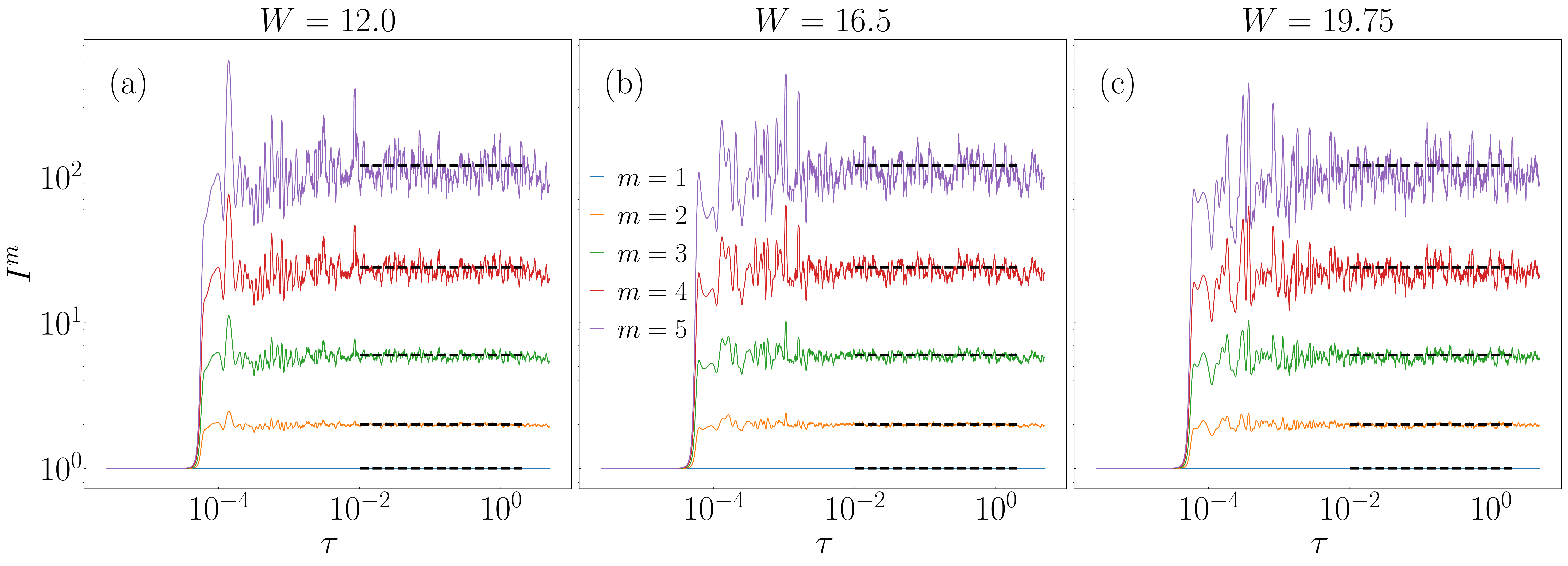}
\vspace{-0.1cm}
\caption{
Time resolved fluctuations $I_m(t)$ of the SFF $K(\tau)$ as defined in Eq.~(\ref{def_Im}).
Results in panels (a)-(c) correspond to the disorders below ($W=12$), at ($W=16.5 \approx W_{\rm c}$) and above ($W=19.75$) the localization transition for the system size $L=40.$ Horizontal lines denote $\tilde I_m = m!$, see Eq.~(\ref{def_iid_phases}). 
}
\label{fig_sff3}
\end{figure}

We conclude this section by a comment on the fluctuations of the SFF $K(\tau)$.
We define higher moments of the $K(\tau)$ as~\cite{flack_bertini_20, chan_deluca_21}
\begin{equation} \label{def_Ktm}
K_m(\tau) = \frac{1}{Z} \left\langle \left|\sum_{\alpha = 1}^V \rho(\varepsilon_\alpha) e^{-i 2\pi\varepsilon_\alpha \tau}\right|^{2m} \right\rangle_W \, ,
\end{equation}
and study the time-resolved ratio of the $m$-th moment relative to the two-point SFF to the power $m$,
\begin{equation} \label{def_Im}
    I_m(t) = \frac{K_m(\tau)}{K(\tau)^m} \,.
\end{equation}

The main question that we address is the sensitivity of the fluctuation measure $I_m(t)$ to the localization transition.
An analytical estimate of $K_m(\tau)$ in the localized phase can be obtained by assuming that the phases $e^{-i2\pi \varepsilon_\alpha \tau}$ are i.i.d.~complex random numbers with a vanishing mean and unit variance.
Neglecting the filter function $\rho(\varepsilon_\alpha)$ in Eq.~(\ref{def_Ktm}), one observes that
\begin{align}
    & \left\langle\left|\sum_{\alpha=1}^V e^{-i2\pi \varepsilon_\alpha \tau}\right|^{2m}\right\rangle_W = 
    \left\langle \sum_{\alpha_1, ..., \alpha_m=1}^V \sum_{\beta_1,...,\beta_m=1}^V
    e^{-i2\pi(\varepsilon_{\alpha_1}+...+\varepsilon_{\alpha_m} - \varepsilon_{\beta_1} - ... - \varepsilon_{\beta_m})\tau}\right\rangle_W \nonumber \\
  \approx & \sum_{\alpha_1, ..., \alpha_m=1}^V \sum_{\beta_1,...,\beta_m=1}^V \, \sum_{\Pi \in S_m} \delta_{(\alpha_1,...,\alpha_m),(\beta_{\Pi_1},...,\beta_{\Pi_m})} = V^m \, m! \,, \label{def_iid_phases}
\end{align}
where $S_m$ denotes the symmetry group of $m$ objects, which contains $m!$ elements.
As a consequence, the fluctuation measure in this approximation equals $\tilde I_m = m!$.
Exactly the same result $I_m=m!$ applies also in the standard RMT ensembles, e.g., the GOE~\cite{flack_bertini_20}.

Numerical results for $I_m(\tau)$ are shown in Fig.~\ref{fig_sff3} for three different disorder values which correspond to the disorder below, at and above the transition.
The results appear to be very similar.
In particular, the prediction $\tilde I_m = m!$, see the horizontal lines in Fig.~\ref{fig_sff3}, represents a good estimate for $I_m(\tau)$ at small $m$.
While the agreement is expected for the GOE regime and the localized regime, we observe no qualitative change in behaviour at the localization transition.
The temporal fluctuations of the SFF $K(\tau)$ hence appear to be insensitive to the localization transition.

\section{Scaling solutions across the transition} \label{sec:scaling}

While the goal of the previous section was to compare different spectrum related methods to pinpoint the critical point, we here turn our attention to numerical scaling solutions of transition indicators across the localization transition.
We first review some important steps of the conventional scaling analyses in Sec.~\ref{sec:scaling_variables}.
The advantage of those scaling analyses, studied extensively in the last decades~\cite{slevin_ohtsuki_99, slevin_ohtsuki_00, milde_roemer_00, slevin_markos_01, rodriguez_vasquez_10, rodriguez_vasquez_11, slevin_ohtsuki_14, Tarquini2017b}, is to provide information about the critical point, the critical exponents, and the corrections to scaling~\cite{slevin_ohtsuki_99}.
However, such procedure usually also requires the scaling functions to be expanded in power series in the vicinity of the critical point.
In Sec.~\ref{sec:floating} we approach the problem from a different perspective. Disregarding our previous knowledge of the problem, we assume the precise location of the critical point in the 3D Anderson model is not known and hence the series expansion of the scaling functions cannot be applied. What other approaches could then be used in order to obtain meaningful information about the critical parameters?
We show that the cost function minimization procedure with a floating crossing point~\cite{suntajs_bonca_20b} represents one of such tools.
It allows for finding the optimal scaling collapse in the targeted parameter region without assuming any particular form of the scaling function, apart from the choice of the family of the correlation length functions.
Moreover, it also allows for detecting the finite-size corrections to the critical point, which in the case of the 3D Anderson model share some similarities with the irrelevant scaling variables used in the context of conventional scaling analyses.

\subsection{Scaling analysis using relevant and irrelevant scaling variables} \label{sec:scaling_variables}

Let us consider a dimensionless transition indicator $\Lambda$, for which we apply the scaling analysis framework used by Slevin and Ohtsuki~\cite{slevin_ohtsuki_99}.
In their work $\Lambda$ represented a scaled correlation length, while here we refer to it as a general localization transition indicator.
Then, building on the renormalization group arguments for the scaling behavior across the transition~\cite{cardy_96}, $\Lambda$ is expected to behave in finite systems as
\begin{equation} \label{def_scaling_function}
\Lambda = F(\chi L^{1/\nu}, \psi L^y) \,,
\end{equation}
where $F$ represents some unknown function, and $\chi$ and $\psi$ are scaling variables.
They are both functions of the scaled disorder, $\chi = \chi(w)$ and $\psi=\psi(w)$, where $w = (W - W_c)/W_c$ and $W_c$ is the critical point.
We refer to $\chi$ as a relevant scaling variable and $\psi$ is the dominant irrelevant scaling variable.
The latter is irrelevant since we assume $y<0$ in Eq.~(\ref{def_scaling_function}), i.e., its contribution is negligible for large enough $L$.

A standard assumption is that the irrelevant contribution is small and hence the function $F$ in Eq.~(\ref{def_scaling_function}) can be expanded in a power series around $\psi L^{y} = 0$, yielding
\begin{equation} \label{def_ni_expansion}
\Lambda = \sum_{\alpha=0}^{n_I} (\psi L^{y})^\alpha F_\alpha(\chi L^{1/\nu}) \,,
\end{equation}
where $F_\alpha(\chi L^{1/\nu}) \equiv \frac{1}{\alpha!} \frac{\partial^\alpha F(\chi L^{1/\nu}, \psi L^y)}{\partial (\psi L^y)^\alpha} |_{\psi L^y = 0}$.
Since in general none of the functions $F_\alpha(\chi L^{1/\nu})$ in Eq.~(\ref{def_ni_expansion}) are known, it is convenient to expand them in a power series around the critical point,
\begin{equation} \label{def_nr_expansion}
F_\alpha (\chi L^{1/\nu}) = \sum_{\beta=0}^{n_R} (\chi L^{1/\nu})^\beta F_{\alpha\beta} \,,
\end{equation}
where $F_{\alpha\beta} = \frac{1}{\beta!} \frac{\partial^\beta F_\alpha(\chi L^{1/\nu})}{\partial (\chi L^{1/\nu})^\beta} |_{\chi L^{1/\nu}=0}$.
Moreover, the scaling variables can also be expanded in power series,
$\chi(w) = \sum_{n=1}^{m_R} b_n w^n$ and
$\psi(w) = \sum_{n=0}^{m_I} c_n w^n$.
This leads in total to $N_{\rm fit} = (n_I+1)(n_R+1)+m_I+m_R+2$ free fitting parameters.
It is clear that such expansions are efficient if the critical point can be located to a sufficient accuracy without the scaling analysis.
Indeed, this seems to be a reasonable assumption for the 3D Anderson model.

The usual choice made by the majority of works including ours, is that the relevant scaling variable is linear, $\chi(w) \propto w$.
We hence set $m_R = 1$ further on.
Still, there are plenty of other options for the truncation of the series introduced above.
For example, in the works by Slevin, Ohtsuki and collaborators, they tested robustness of critical parameters towards the choice of the cutoff parameters $n_{\rm R}, n_{\rm I}, m_{\rm I}$ to describe the scaling properties of the dimensionless correlation length~\cite{slevin_ohtsuki_99, slevin_ohtsuki_00, slevin_ohtsuki_14} and the wave function multifractality~\cite{rodriguez_vasquez_10, rodriguez_vasquez_11}.
Recently, Tarquini, Biroli and Tarzia~\cite{Tarquini2017b} applied a similar scaling analysis using $n_{\rm I} = 1$ and $m_{\rm I} = 0$ to seek for the scaling collapse of the level spacing ratio $r$ from Eq.~(\ref{eq:r_ratio}).
The latter quantity is also going to be the focus of our analysis in the next section.

\subsection{Scaling analysis using the cost function minimization with a floating crossing point} \label{sec:floating}

In contrast to the analysis in the previous section, there are many systems in which the critical point in the thermodynamic limit is not known to sufficient accuracy.
In such case the low-order polynomial expansion of the scaling function, as expressed in Eqs.~(\ref{def_ni_expansion}) and~(\ref{def_nr_expansion}), may be inefficient.

The cost function minimization approach, introduced in~\cite{suntajs_bonca_20b} and applied in~\cite{suntajs_bonca_20a}, is a numerical approach to find the optimal scaling collapse of a transition indicator $\Lambda$ as a function of $L/\xi$,
where $\xi$ is the correlation length.
Let $\Lambda$ be a function of the disorder $W$ and system size $L$.
Then, we refer to the optimal scaling collapse as the one that minimizes the cost function ${\cal C}_\Lambda$,
\begin{equation}
    {\cal C}_\Lambda = \frac{\sum_{j=1}^{N_{\rm p}-1}|\Lambda_{j+1}-\Lambda_j|}{{\rm max}\{\Lambda_j\} - {\rm min}\{\Lambda_j\}} \,,
\end{equation}
where $N_{\rm p}$ denotes the number of all data points of $\Lambda(W,L)$ included in the cost function minimization procedure, and $\Lambda_j$ denotes the value of $\Lambda$ at a given $W$ and $L$.
We sort all $N_{\rm p}$ values of $\Lambda_j$ according to nondecreasing values of ${\rm sign}[W-W^*] L/\xi$, where $W^*$ is referred to as the crossing point in finite systems that eventually converges to the critical point $W_{\rm c}$ in the thermodynamic limit $L \to \infty$.
Then, using a differential evolution method, we obtain the best data collapse by finding the fitting parameters for which the cost function ${\cal C}_\Lambda$ is minimal. 
The cost function is defined such that the ideal data collapse yields ${\cal C}_\Lambda = 0$, and ${\cal C}_\Lambda > 0$ otherwise.

One of the main advantages of the cost function approach is that it does not require any knowledge about the scaling function.
Therefore, its application is not restricted to the data in the vicinity of the critical point, where the scaling function can be approximated using series expansion.
Nevertheless, a convenient aspect of the conventional scaling analysis discussed in Sec.~\ref{sec:scaling_variables} is that it accounts for the drift of the crossing point in finite systems due to the presence of irrelevant scaling variables.
We mimic this effect by allowing for the crossing point $W^*$ to be system size dependent, i.e., $W^* = W^*(L)$.
The crossing point $W^*(L)$ enters the scaling analysis through the linear scaling variable $\chi = W-W^*(L)$.
We then denote the scaling variable as $\chi = \chi(W,L)$, and the scaling function can be written as
\begin{equation} \label{scaling_floating}
\Lambda = G[\chi(W,L) L^{1/\nu}] \,.
\end{equation}
In this approach the optimal scaling solution may shift the curves $\Lambda(W)$ at a fixed $L$ in the horizontal direction, but not in the vertical direction.
Its application is therefore convenient when seeking for global scaling collapses of quantities that have well defined upper and lower values.
We dub the method as {\it the cost function minimization approach with a floating crossing point}.

Below we test our approach using the level spacing ratio $r$ defined in Eq.~(\ref{eq:r_ratio}).
Its simplest application is shown in Fig.~\ref{fig_r_zoom}.
The search of the numerical scaling solution is built on the preexisting knowledge of the location of the critical point, since the data included in the cost function minimization procedure correspond to a narrow symmetric interval around $W_{\rm c}$, i.e., $W \in [W_{\rm c} - 1, W_{\rm c} + 1]$.
In this case, it is sufficient to consider the crossing point $W^*(L)$ as a system independent parameter $w_0$.
Then, the function $G$ in Eq.~(\ref{scaling_floating}) is a function of $L/\xi$, where $\xi = |W-w_0|^{-\nu}$.
Using the cost function minimization procedure we find the optimal values of two fitting parameters $w_0 = 16.53$ and $\nu=1.49$, which are very close to the critical parameters given in Eq.~(\ref{def_Wc}).

\begin{figure}[!t]
\includegraphics[width=1.0\columnwidth]{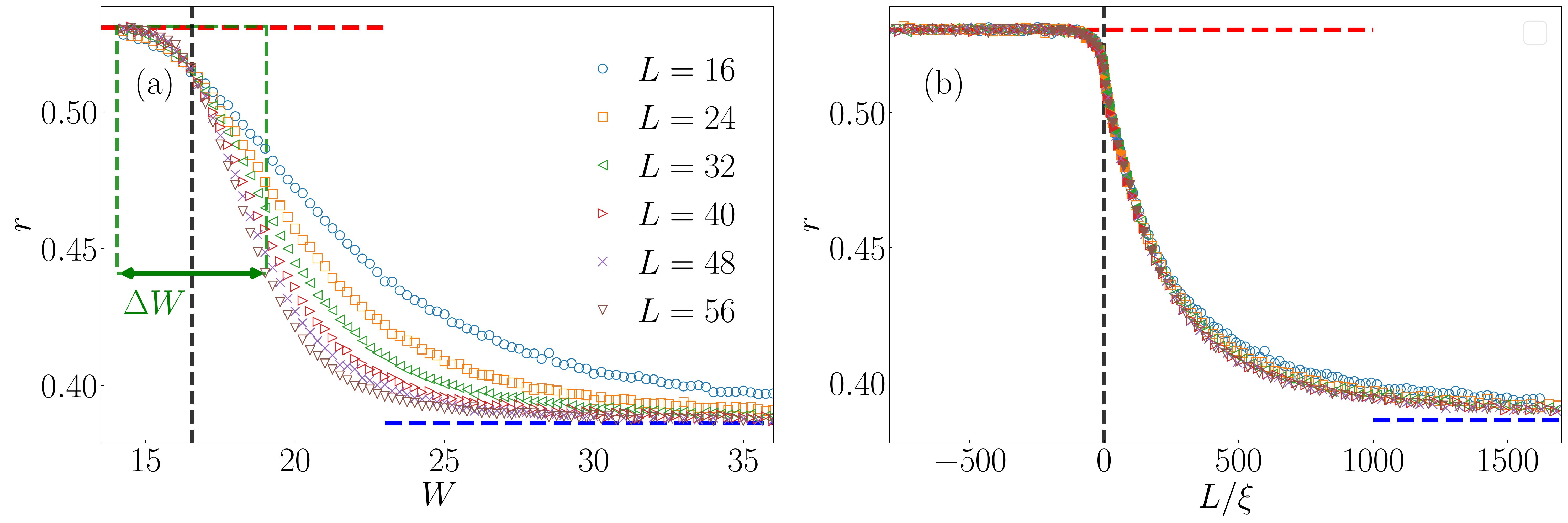}
\vspace{-0.1cm}
\caption{
Numerical scaling solution of the average level spacing ratio $r$ from Eq.~(\ref{eq:r_ratio}).
Each data point was obtained by performing the average over $N_\mathrm{samples}=1500$ disorder realizations.
The step size between consecutive data points equals $\delta W=0.25$ in units of $W$.
The upper and lower dashed horizontal lines denote the GOE and Poisson values $r_{\rm GOE} \approx 0.5307$ and $r_{\rm PE} \approx 0.3863$, respectively.
(a)
Dependence of $r$ on the disorder strength $W$ for different system sizes $L$.
The vertical dashed line denotes the critical point $W_{\rm c} = 16.54$.
The green rectangular frame denotes the data points that were provided as an input to the cost function minimization algorithm with $\Delta W$ denoting the width of the interval.
(b)
Scaling collapse of the data from the left panel as a function of $L/\xi$, where $\xi = {\rm sign}[W-w_0] \times |W-w_0|^{-\nu}$.
Filled symbols denote the data points from the interval $\Delta W$ (the green rectangular frame in the left panel) which were used in the actual scaling collapse.
The parameters of the optimal scaling solution are $w_0=16.56$ and $\nu=1.46$.
}
\label{fig_r_wide}
\end{figure}

In the remainder of the analysis we focus on two main goals.
The first goal is to study the role of the target window for which the scaling solution is obtained.
As discussed in Sec.~\ref{sec:measures} and demonstrated in Fig.~\ref{fig_r_wide}(a), $r$ extends from the GOE regime below the transition with $r_{\rm GOE} \approx 0.5307$, to the Poisson regime above the transition with $r_{\rm PE} \approx 0.3863$.
The question that we address is: supposing that the entire crossover region of $r$ is included in the search of the scaling solution, can one still get an accurate estimate of the critical point?
The second goal is to test how accurately the critical point can be estimated if one leaves the crossing point in finite systems to float with system size, i.e., the crossing point $W^*(L)$ may be an arbitrary function of the system size.
The use of such floating crossing point is very convenient if the accurate location of the critical point is not known and results in a wide disorder interval are included in the search of the scaling solution.

\begin{figure}[!t]
\includegraphics[width=1.0\columnwidth]{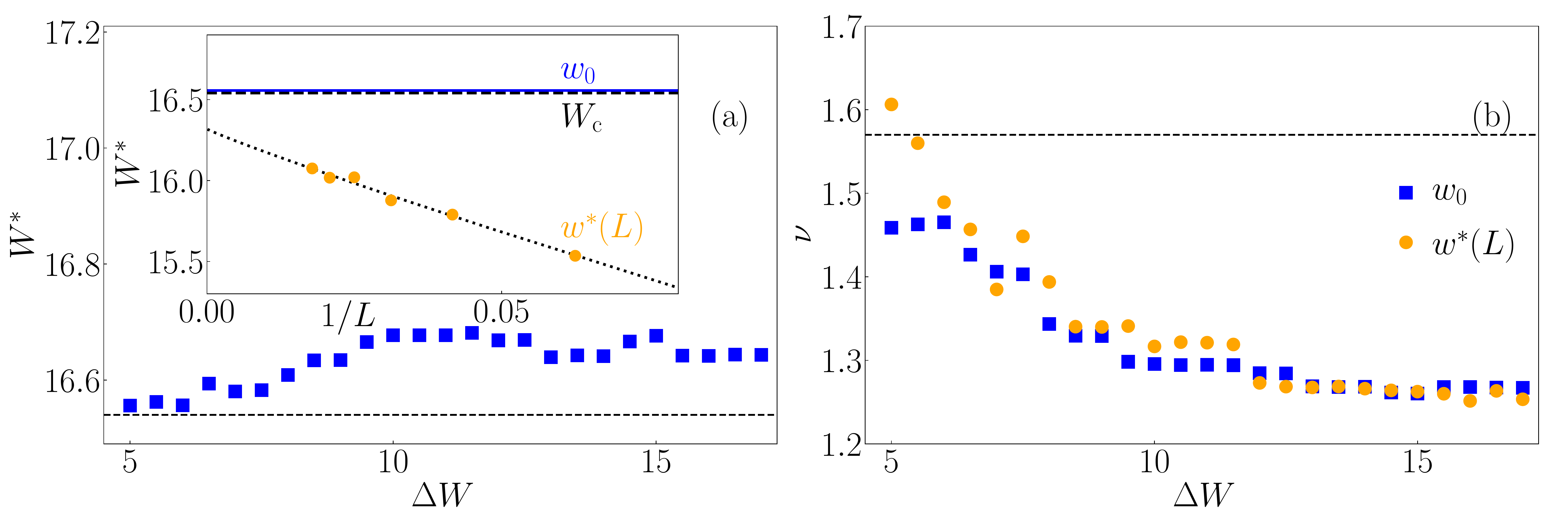} 
\vspace{-0.1cm}
\caption{
Parameters of the optimal scaling solutions of the average level spacing ratio $r$.
In the cost function minimization procedure, we include the data in Fig.~\ref{fig_r_wide}(a) from some interval of width $\Delta W$.
Specifically, we include the results for the disorders $W$ such that $W \in [W_{\rm min},W_{\rm min} + \Delta W]$, where $W_{\rm min} = 14.04$.
Squares denote results for the case when the crossing point is a system size independent constant $W^* = w_0$:
(a) $W^*$ versus $\Delta W$, (b) $\nu$ versus $\Delta W$.
Horizontal lines denote the corresponding critical values from Eq.~(\ref{def_Wc}).
The inset of (a) studies the optimal scaling solution for $\Delta W = 5$ when the crossing point $W^*(L)$ is an independent fitting parameter for each system size.
The dotted line is a three-parameter fit of $W^*(L)$ to the function $w^* + w_1 L^{\zeta}$, where $w^* = 16.32$ and $\zeta \approx -0.93$.
Dashed and solid horizontal lines correspond to $W_{\rm c}$ from Eq.~(\ref{def_Wc}) and $w_0$ at $\Delta W = 5$.
}
\label{fig_Wc_eta}
\end{figure}

The rectangular frame in Fig.~\ref{fig_r_wide}(a) is a sketch to mark the data which are included in the search of the optimal scaling solution.
The width of the frame is denoted by $\Delta W$.
To obtain the scaling collapse in Fig.~\ref{fig_r_wide}(b) we use $\Delta W$ as marked in Fig.~\ref{fig_r_wide}(a), while the analysis in Fig.~\ref{fig_Wc_eta} is carried out as a function of varying $\Delta W$.
The scaling solution in Fig.~\ref{fig_r_wide}(b) is obtained using an identical cost function minimization procedure as in Fig.~\ref{fig_r_zoom}(b) since in both cases there are two free fitting parameters, the crossing point $W^* = w_0$ and the critical exponent $\nu$.
However, since in Fig.~\ref{fig_r_wide} we use a larger $\Delta W$ than in Fig.~\ref{fig_r_zoom}, as well as a larger step between consecutive data points and less disorder averaging, the resulting optimal parameters (see figure captions for details) are slightly less accurate when compared to the values in Eq.~(\ref{def_Wc}).

Next we increase $\Delta W$ and study how the critical parameters evolve as a function of $\Delta W$.
We first consider the simplest case where the crossing point $W^*$ is a system size independent number $w_0$.
We show $w_0$ versus $\Delta W$ in Fig.~\ref{fig_Wc_eta}(a) and $\nu$ versus $\Delta W$ in Fig.~\ref{fig_Wc_eta}(b).
It is interesting to compare the results at relatively small $\Delta W \approx 5$ (also used in Fig.~\ref{fig_r_wide}) and at large $\Delta W \approx 15$.
In the latter case the data included in the cost function minimization extend over almost the entire crossover region between $r_{\rm GOE}$ and $r_{\rm PE}$.
We observe that the crossing point is determined very accurately even at $\Delta W \approx 15$.
In contrast, the optimal value of $\nu$ deviates much stronger at larger $\Delta W$.
These results suggest a considerable robustness of the critical point $W_c$ towards the finite-size effects.
This observation seems to be related to the fact that even numerical studies of the 3D Anderson model in the seventies and early eighties, using very limited computational resources, detected the critical properties with reasonable accuracy~\cite{prelovsek_79, weaire_kramer_80, B.KramerA.MacKinnon1981}.

Finally we apply the cost function minimization procedure to the data in the rectangular frame in Fig.~\ref{fig_r_wide}(a) using the free crossing point $W^*(L) = w^*(L)$.
In this case, the crossing point is an independent fitting parameter $w^*(L)$ for each $L$.
Hence the total number of fitting parameters is seven (six parameters $w^*(L)$ for six different system sizes, and the critical exponent $\nu$).
The size dependent crossing point $W^*(L)$ is shown in the inset of Fig.~\ref{fig_Wc_eta}(a).
It exhibits two important features.
First, when extrapolated to $1/L \to 0$, it agrees with the value of the critical point $W_{\rm c}$ in Eq.~(\ref{def_Wc}) rather accurately.
Second, we find a good fit for $W^*(L)$ with $W^*(L) = w^* + w_1 L^{\zeta}$, where $\zeta \approx -0.93$.
This is close to the recent study of the scaling properties of $r$ using the irrelevant scaling parameter $\propto L^{y}$, which reported $y \approx -1$~\cite{Tarquini2017b}.
We note, though, that the latter agreement is a mere observation since we are not aware of a theory predicting that $\zeta$ and $y$ should be identical.

The results of this section suggest two convenient properties of the cost function minimization approach with a floating crossing point: it can be applied to systems in which the critical point is not yet known precisely, and it mimics to a certain extent the role of irrelevant scaling variables.
For the specific example of the 3D Anderson model, the floating crossing point in finite system convincingly converged to the vicinity of the critical point predicted by other methods.

\section{Conclusions}

In this paper we discussed, compared and calculated several characteristic spectral features of the 3D Anderson model.
While all those properties were to some extent already studied in the past, we considered them through the lens of modern numerical approaches and put an effort into making quantitative connections between them.
The main focus was devoted into the comparison between the level sensitivity statistics and the level statistics.
Specifically, we compared the width of the level curvatures distribution and the Thouless time extracted from the spectral form factor.
We showed that, using proper normalization, both quantities behave very similarly and provide an efficient tool to pinpoint the localization transition.
We then also made a quantitative connection of those quantities to the quantities studied in the work by Edwards and Thouless~\cite{edwards_thouless_72}.
Finally, we study the numerical scaling solutions of the level spacing ratio $r$ in the vicinity of the critical point.
We showed that the cost function minimization approach with a floating crossing point~\cite{suntajs_bonca_20b} provides a framework to accurately pinpoint the critical point.
Moreover, it provides an additional information about the finite-size corrections that to certain extent resemble the behavior of irrelevant scaling variables used in the conventional scaling analysis.

\section{Acknowledgments}

We acknowledge insightful discussions with P. Prelov\v sek.
This work was supported by the Slovenian Research Agency (ARRS), Research Core Fundings Grants No.~P1-0044, No.~P1-0402 and No.~J1-1696, and by European Research Council (ERC) under Advanced Grant 694544 -- OMNES.

\bibliography{mybibfile,Anderson_localization,PhD-MBL,refs_Anderson}


\end{document}